\documentclass[twocolumn]{aastex62}

\usepackage{longtable} 

\received{...}
\revised{...}
\accepted{...}
\submitjournal{ApJ}

\shorttitle{HXR emission in flare-related coronal jets}
\shortauthors{Musset et al.}

\begin{document}

\title{Statistical study of hard X-ray emitting electrons associated with flare-related coronal jets}

\correspondingauthor{Sophie Musset}
\email{smusset@umn.edu}

\author{Sophie Musset}
\affil{School of Physics and Astronomy \\
University of Minnesota \\
116 Church St SE \\
Minneapolis, MN 55455, USA}

\author{Mariana Jeunon}
\affiliation{School of Physics and Astronomy \\
University of Minnesota \\
116 Church St SE \\
Minneapolis, MN 55455, USA}

\author{Lindsay Glesener}
\affiliation{School of Physics and Astronomy \\
University of Minnesota \\
116 Church St SE \\
Minneapolis, MN 55455, USA}



\begin{abstract}


We present the statistical analysis of 33 flare-related coronal jets, and discuss the link between the jet and the flare properties in these events. 
We selected jets that were observed between 2010 and 2016 by the Atmospheric Imaging Assembly (AIA) on board the Solar Dynamic Observatory (SDO) and are temporally and spatially associated with flares observed by the Reuven Ramaty High Energy Solar Spectrometric Imager (RHESSI). For each jet, we calculated the jet duration and projected velocity in the plane of sky. The jet duration distribution has a median of 18.8 minutes. The projected velocities are between 31 km/s and 456 km/s with a median at 210 km/s. For each associated flare, we performed X-ray imaging and spectroscopy and identify non-thermal emission.
Non-thermal emission was detected in only 1/4 of the events considered. We did not find a clear correlation between the flare thermal energy or SXR peak flux and the jet velocity. A moderate anti-correlation was found between the jet duration and the flare SXR peak flux. There is no preferential time delay between the flare and the jet. The X-ray emission is generally located at the base of the jet. The analysis presented in this paper suggests that the flare and jet are part of the same explosive event, that the jet is driven by the propagation of an Alfvenic perturbation, and that the energy partition between flare and jets varies substantially from one event to another.

\end{abstract}




\section{Introduction} \label{sec:intro}


How energetic particles are accelerated in the corona and how they escape the Sun's atmosphere to be detected in the interplanetary medium are still outstanding questions in solar physics. The most direct diagnostic of energetic electrons in flares is the bremsstrahlung emission produced as they interact with the ambient dense plasma of the Sun's low corona and chromosphere, emitted in hard X-rays (HXR). Energetic electrons are accelerated during flares, when magnetic energy is suddenly released. While the energy budget of solar flares is still under investigation, observations suggests that a large fraction of the released energy (20\% to 50\%) is transmitted to particle acceleration \cite[see e.g.][]{emslie2004,emslie2005,emslie_etal_2012}.
A distinction is generally made between confined flares, for which there is no evident ejection of plasma nor energetic particles escaping the Sun's atmosphere, and eruptive flares, for which the magnetic energy release can be accompanied with plasma ejections (filament eruptions and/or coronal mass ejections, CMEs). Some flares are associated with energetic particles that can be then detected in the interplanetary medium. 
Evidence of escaping energetic electron beams can also be detected in the high corona and in the interplanetary medium as they disturb the ambient plasma, creating Langmuir waves that decay into electromagnetic waves, observed as type III radio bursts ( see \citealt{reid_2014_review} for a review of type III radio bursts and \citealt{reid_vilmer_2017} for a study of the link between coronal type III bursts and the associated X-ray flares). 
While such emissions are regularly observed, even in non-eruptive events, the context in which energetic electrons accelerated in the low corona gain access to the interplanetary medium is still not fully understood. 
To escape the solar corona, energetic electrons have to gain access to open magnetic field lines. Such open field lines are sometimes illuminated by plasma ejections, such as coronal jets.


Solar X-ray jets have been first detected in soft X-rays by the Soft X-ray Telescope (SXT) on board Yohkoh \citep[see e.g.][]{shibata_etal_1989}, and are described as collimated ejections of plasma. Hundred of jets have been observed in the SXT era, and jets have also been detected in extreme ultraviolet (EUV) and ultraviolet \cite[see][for a review of coronal jets]{raouafi_etal_2016}. 
Coronal jets are ubiquitous as they are found in different regions of the solar atmosphere (coronal holes, quiet sun, active regions) and are observed with many spatial and time scales. 


Jets are believed to arise from interchange magnetic reconnection between closed and open magnetic field lines. 
The Shibata interchange jet model \citep{shibata_etal_1992b, yokoyama_shibata_1996} predicts a hot (several MK) jet resulting from a magnetohydrodynamic (MDH) shock produced near the reconnection site, along side a cooler jet resulting from chromospheric evaporation following the rapid energy release during magnetic reconnection. 
The triggering processes for such interchange reconnection and jet generation are discussed in emerging flux reconnection models \citep[see e.g.][]{shibata_etal_1992b, yokoyama_shibata_1996}, 3D reconnection models \citep[see e.g.][]{pariat_etal_2010, pariat_etal_2015, pariat_etal_2016}, and small-scale filament eruption models \citep{nistico_etal_2009, moore_etal_2010, raouafi_etal_2010, sterling_etal_2015, sterling_etal_2016, wyper_etal_2017, wyper_etal_2018}. 
All these models invoke reconnection of stressed magnetic field lines with open magnetic field lines, enabling plasma ejection on those open lines, which is compatible with the interchange reconnection model and with particle escape. \cite{pariat_etal_2016} MHD simulations showed that jet-like events are driven by propagating Alfvenic waves. In a low-$\beta$ plasma, the velocity of the wave is close to the ambiant Alfven speed, much higher than the bluk flow speed of the plasma. Interestingly, \cite{matsui_etal_2012} used the spectroscopic observations of a jet to calculate velocities in different lines and showed that the velocity of the cool plasma was greater than the upper limit of the velocity expected from chromospheric evaporation, suggesting that the cooler plasma revealed the velocity of the MHD waves driving the jet.
 
The idea that all jets could be small-scale filament eruptions (mini-CMEs), where the filament could not always be resolved by the current EUV instruments, implies that there would be some sort of scaling between the amount of energy released and the size, speed and morphology of the jet. For flare-related jets, this could potentially mean that jet size and speed scales with the intensity of the flare, since there is a correlation between the flare intensity (e.g. it's GOES class) and the CME energy \citep[see e.g.][]{emslie_etal_2012}.

While observational and simulation studies have focused on the plasma observation and MHD modeling of jets, there is also evidence for particle acceleration and escape from the solar atmosphere in flare-related jets. 


Solar jets have been associated with energetic electron events observed at 1 AU \cite[e.g.][]{krucker_etal_2011} as well as solar energetic particles \citep[SEPs, e.g.][]{nitta_etal_2008}. When X-ray or extreme ultraviolet (EUV) observations are available for the studied events, these observations show that electron acceleration is associated with coronal jets, and that energetic electrons have access to open magnetic field lines to propagate in the heliosphere. A link between coronal jets and escaping beams of electrons has also been reported using radio diagnostics: several observations of simultaneous type III radio bursts have been reported for X-ray jets \citep{kundu_etal_1995, raulin_etal_1996} and EUV jets \citep{christe_etal_2008, glesener_etal_2012, chen_etal_2013}.
 \cite{glesener_etal_2012} used the imaging capability of the Nancay Radioheliograph (NRH) to study the location of a type III radio burst emission occurring at the time of a EUV jet. They show that the type III burst emission is spatially linked to the open magnetic field line enlightened by the hot thermal emission from the jet. \cite{chen_etal_2013} showed a close relationship between an EUV jet and the decametric type III bursts tracing escaping electron beams, imaged in the low corona with the Very Large Array. 

Hard X-ray (HXR) bremsstrahlung emission, from energetic electrons interacting in the low corona, has also been found to be related to EUV jets. \cite{krucker_etal_2011} looked at the link between jets and energetic electrons for 16 energetic electron events observed at 1 AU, focusing on HXR emission from magnetic loop footpoints ; for 6 of the events, EUV observations were available and coronal jets were detected at the flare site. For the other events, the presence of three footpoints was interpreted as a possible signature interchange reconnection that could lead to coronal jets. Coronal non-thermal emissions in HXR are difficult to observe due to the limited dynamic range of current solar X-ray instruments \citep{sainthilaire_etal_2009}. However, \cite{bain_fletcher_2009} and \cite{glesener_etal_2012} studied two EUV jets for which HXR emission was detected in the jet itself, providing the first constraints on energetic electron populations in the coronal jet. 
Recently, \cite{glesener_fleishman_2018} used the combination of EUV, HXR, and radio diagnostics with magnetic field and density modeling of the active region to characterize the energetic electron distributions on both open and closed magnetic field in a flare-associated coronal jet.




While several studies have shown that jets can be linked to electron acceleration and escape from the solar corona, only a few studies have focused on the link between EUV jet properties and X-ray diagnostics of energetic electrons in the low corona. 
It is still unclear what kind of flares are associated with jets, and how the energetic electron distribution is related or not to the jet properties. 
We perform in this paper the first statistical analysis of energetic electrons associated with flare-related jets, during the 8-year coverage of the Reuven Ramaty High Energy Solar Spectroscopic Imager \citep[RHESSI,][]{rhessi} and the Atmospheric Imaging Assembly \citep[AIA,][]{aia} on board the Solar Dynamic Observatory (SDO). 
We identified 33 EUV jets observed by SDO that are associated with a RHESSI flare. The methodology to create this list and analyze the events is described in section \ref{sec:metho}. The statistical analysis of the results is presented in section \ref{sec:analysis} and discussed in section \ref{sec:discussion}.


\section{Methodology} \label{sec:metho}

\subsection{Event selection} \label{sec:selection}

\begin{figure}[t]
\includegraphics[width=\linewidth]{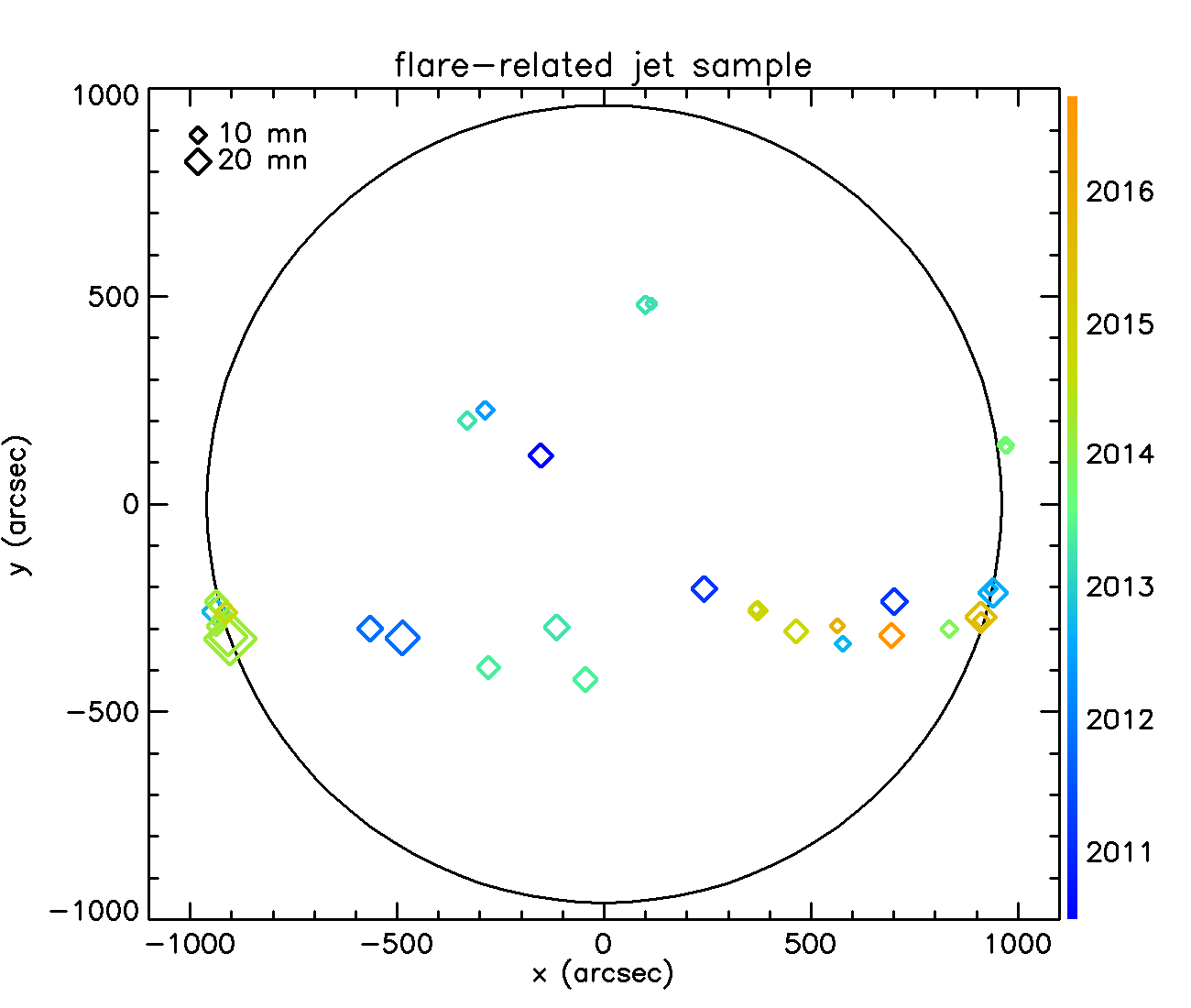}
\caption{Distribution of our jets: position on the solar disk. Date of observation is denoted by the color scale. The size of the symbol is proportional to the duration of each jet. }
\label{jet-sample}
\end{figure}

The flare-related jet events were selected using the list of coronal jets reported in the Heliophysics Events Knowledgebase (HEK) database\footnote{\url{https://www.lmsal.com/hek/}}, and the RHESSI flare list\footnote{\url{https://hesperia.gsfc.nasa.gov/rhessi3/data-access/rhessi-data/flare-list/index.html}}. 
The coronal jets in the HEK have been reported by human users in AIA and IRIS (Interface Region Imaging Spectrograph) EUV observations. 
We used the spatial coordinates of the event, its start and end time, as well as the coordinates of the bounding box, provided in the HEK database.
The RHESSI flare list is automatically computed by comparing the 6-12 keV count rate to a threshold rate, determined from the background level. The flare position reported in the list is calculated for the 6-12 keV energy range and therefore most likely correspond to the position of coronal thermal emission during the flare. We used the flare spatial coordinates as well as the start and end times of the flare.

A preliminary list of candidates was drafted using the following criteria:
\begin{itemize}
\itemsep0em 
\item The time delay between the jet and the flare time intervals is less than 15 minutes
\item The distance between the position of the jet reported in HEK and the position of the flare in the RHESSI list is less than 100 arcsec (i.e. the jet and flare are in the same active region), or the position of the flare is within the bounding box defined for the jet in the HEK database.
\end{itemize}
However, the list of candidates provided by this cross-examination of the two databases had to be refined. Indeed, coronal jet events reported in the HEK database often cover several hours and recurring jet events; and the reported position of jets can be approximate.

Therefore, for each jet candidate in this preliminary list, we looked at the EUV data to perform our own assessment of the time and location of each individual jet. Here, an individual jet is defined as an event for which we see ejection of plasma without discontinuity in time. It may be composed of several shorter individual bursts that cannot be separated by just looking at the ejecta time evolution in the EUV images. At this point, a list of 73 flare-related jets events were visually identified in the EUV 304 \AA \ AIA channel.

We then selected individual events for which a RHESSI flare was found at less than 100 arcsec from the jet position, with a time delay less than 10 minutes between the jet and flare time intervals (less than 10 minutes between the end of the jet and the beginning of the flare if the jet is happening first, and vice-versa). Many jets in our initial list were discarded for one of the following reasons: several flares happening in different active regions at the time of the jet, partial coverage of the flare by the RHESSI observations (due to night or SAA time intervals).
The resulting list contains 33 events, and can be found in appendix \ref{sec:listjets}. It should be noted that since coronal jets in the HEK database are reported by human eye from  UV and EUV observations, jet events can be missing in that list. Therefore, the list created for this paper is not an exhaustive list of jets associated with RHESSI flares, but a sample of such jets. A visual summary of the list is given in figure \ref{jet-sample}.

For each event in that list, we used EUV observations to determine the location of the base of the jet, the jet timing and the jet projected velocity. We used the soft X-ray flux to determine the flare size (GOES class) and hard X-ray observations to study accelerated electrons in the jet-associated flares.

\begin{figure}
\includegraphics[width=\linewidth]{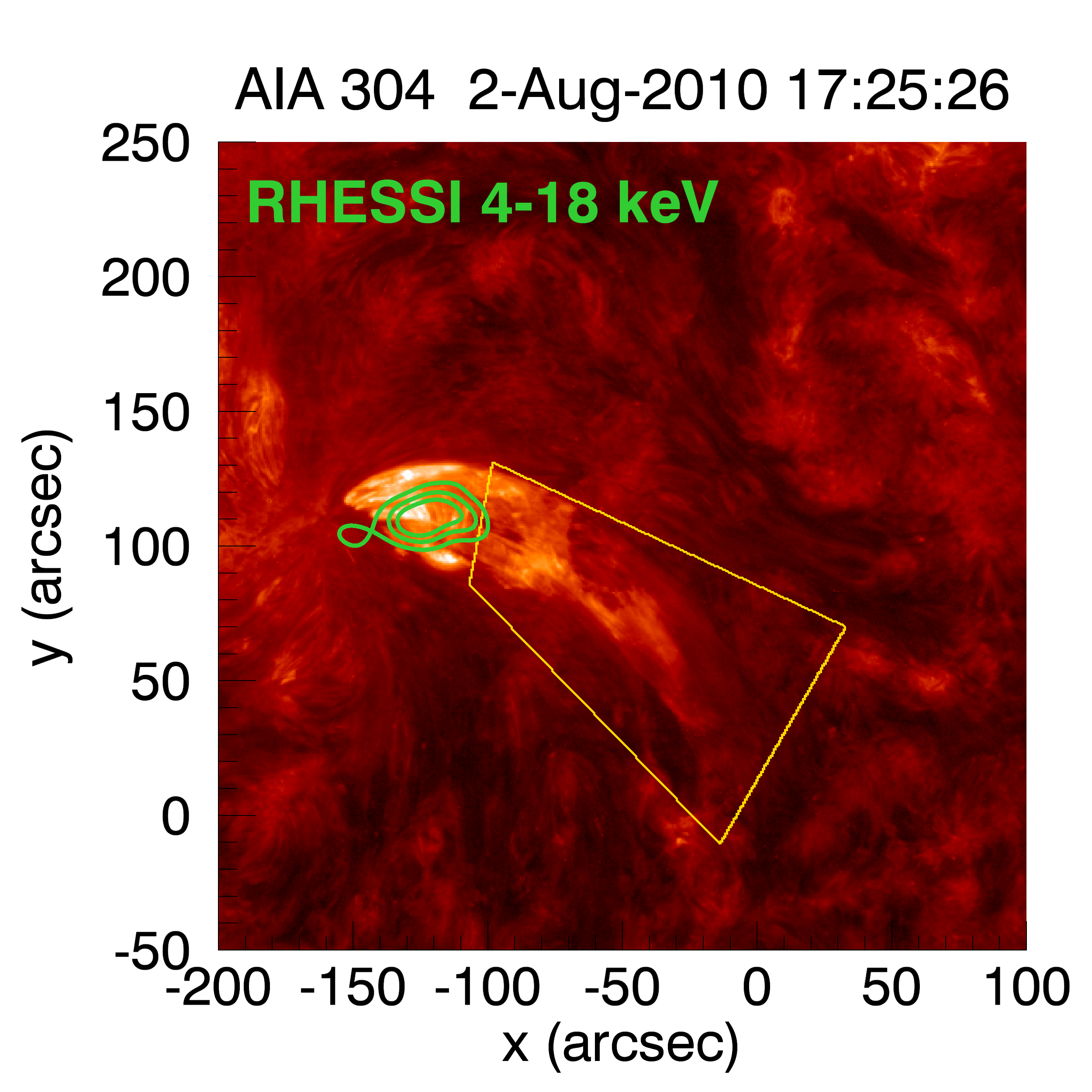}
\caption{Jet on Aug 2 2010 at 304 \AA. White contours are 30 \%, 50 \% and 70 \% contours of the RHESSI CLEAN image in the 4-18 keV energy range, integrated 2 minutes after 17:23:30 UT. Gold box: selected region around the EUV jet. A movie version of this figure is available online. }
\label{roi}
\end{figure}

\subsection{EUV Jet analysis} \label{sec:jets_properties}

The Atmospheric Imaging Assembly (AIA) provides full-Sun imaging in several UV and EUV passbands, with a 12-second cadence and a 1.5 arcsecond spatial resolution \citep{aia}. Launched in 2010, it provides detailed imaging of the chromosphere, transition region and low corona, for plasma temperature of the order of 0.1 to 20 MK. For each of the selected jets, we used the AIA data in order to inspect the time evolution of the morphology, intensity and projected velocity of the jet. 

\subsubsection{Jet timing} \label{sec:jets_timing}

To determine the time evolution of a jet, we computed jet ``lightcurves'' by calculating the time evolution of the mean intensity in the AIA images in a region of interest closely related to the jet. 
Such a region is illustrated in figure \ref{roi} for the 2010 Aug 02 jet. This region was chosen to enclose the jet when most visible in the 304 \AA \ AIA channel, and to avoid contamination by the flare emission. The differential rotation of the Sun was compensated using the SSWIDL routine drot.pro, for jet events located on the solar disk. It should be noted that excluding the base of the jet for the timing analysis can introduce a delay in the computed start time of the jet, of the order of 1 to 2 minutes.

\begin{figure}
\includegraphics[width=\linewidth]{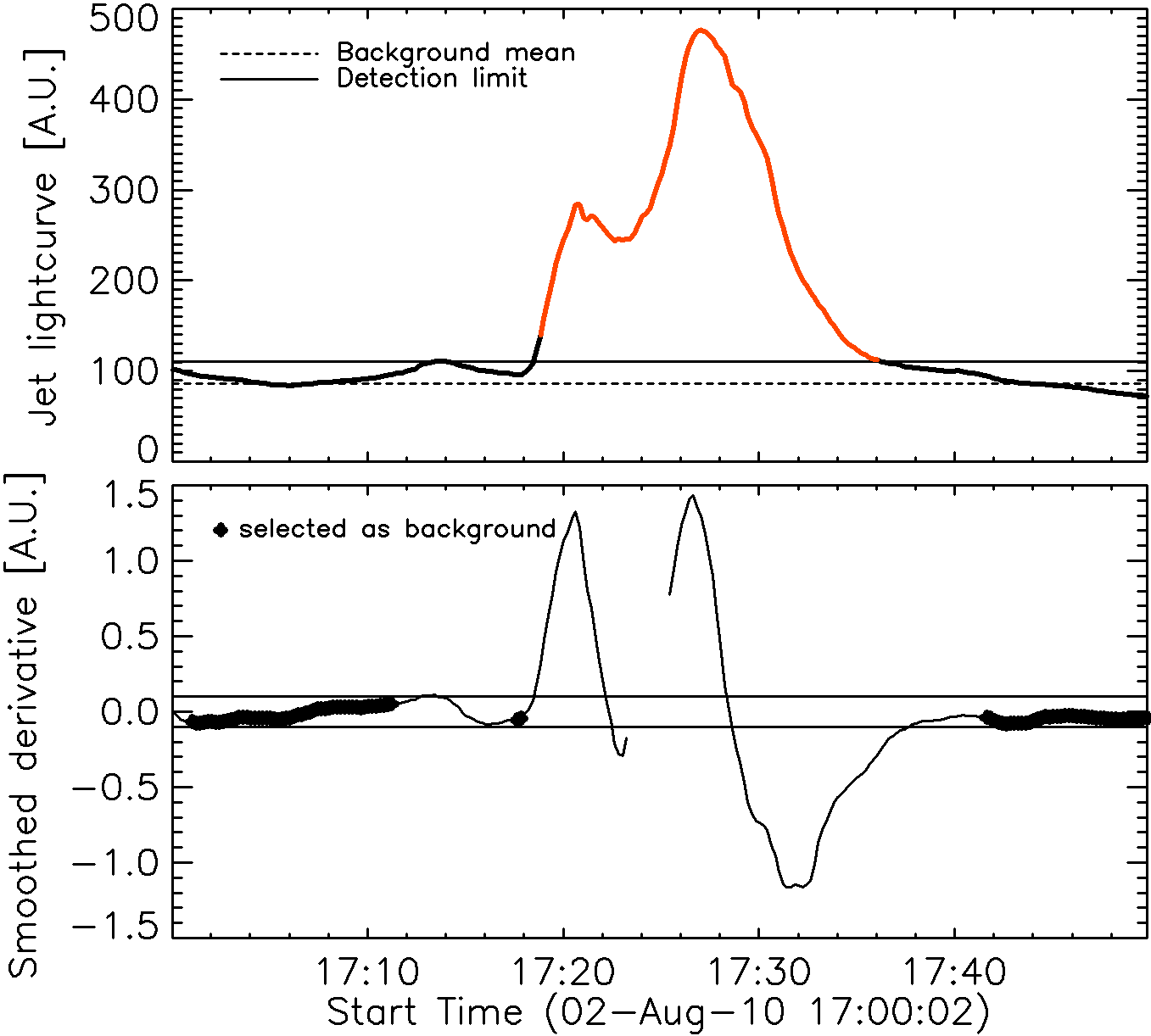}
\caption{Example of the automatic detection of a jet start and end time. Top: mean intensity of the AIA 304 \AA \ signal calculated in the region shown in figure \ref{roi} over time (jet lightcurve). Bottom: smoothed derivative of the jet lightcurve. The horizontal lines in the bottom panel show the region of low derivative used to select background. The selected background interval is shown in bold in the bottom panel. The mean value of the background and the 3$\sigma$ detection limit are shown in the top panel. The red portion of the lightcurve is the portion detected as the jet. The method used to select the background and the jet are described in \ref{sec:jets_timing}. }
\label{jet_duration_ex}
\end{figure}

The starting and ending times of the jet were then automatically calculated using the jet lightcurve with the following procedure:
\begin{enumerate}
\itemsep0em 
\item The derivative of the lightcurve is computed and smoothed using a running mean over 2 minutes (10 data points)
\item The background emission is selected as the part of the lightcurve where the smoothed derivative is close to zero (between -0.1 and 0.1) and where the lightcurve signal is lower than the its average value.
\item The mean and standard deviation of the EUV signal are calculated in the background regions selected.
\item The jet is identified as the time for which the EUV lightcurve exceeds the median background values by three times the standard deviation in the background (3$\sigma$ detection). 
\end{enumerate}
Note that the background can be varying for a sub-sample of jet. In that case, instead of taking the median value of the background signal as the background, a linear regression was performed and used as the time-dependent background value.
The peak time of the jet is also identified during this process, as the position of maximum intensity in the lightcurve for each identified jet.
Figure \ref{jet_duration_ex} shows an example of the background and jet selection, for the 2010 Aug 2 jet. The start and peak time of the jet are reported in the list in appendix \ref{sec:listjets}. These values are used to compute the jet duration that is used for the statistical analysis of jets in this paper.

\begin{figure}
\includegraphics[width=\linewidth]{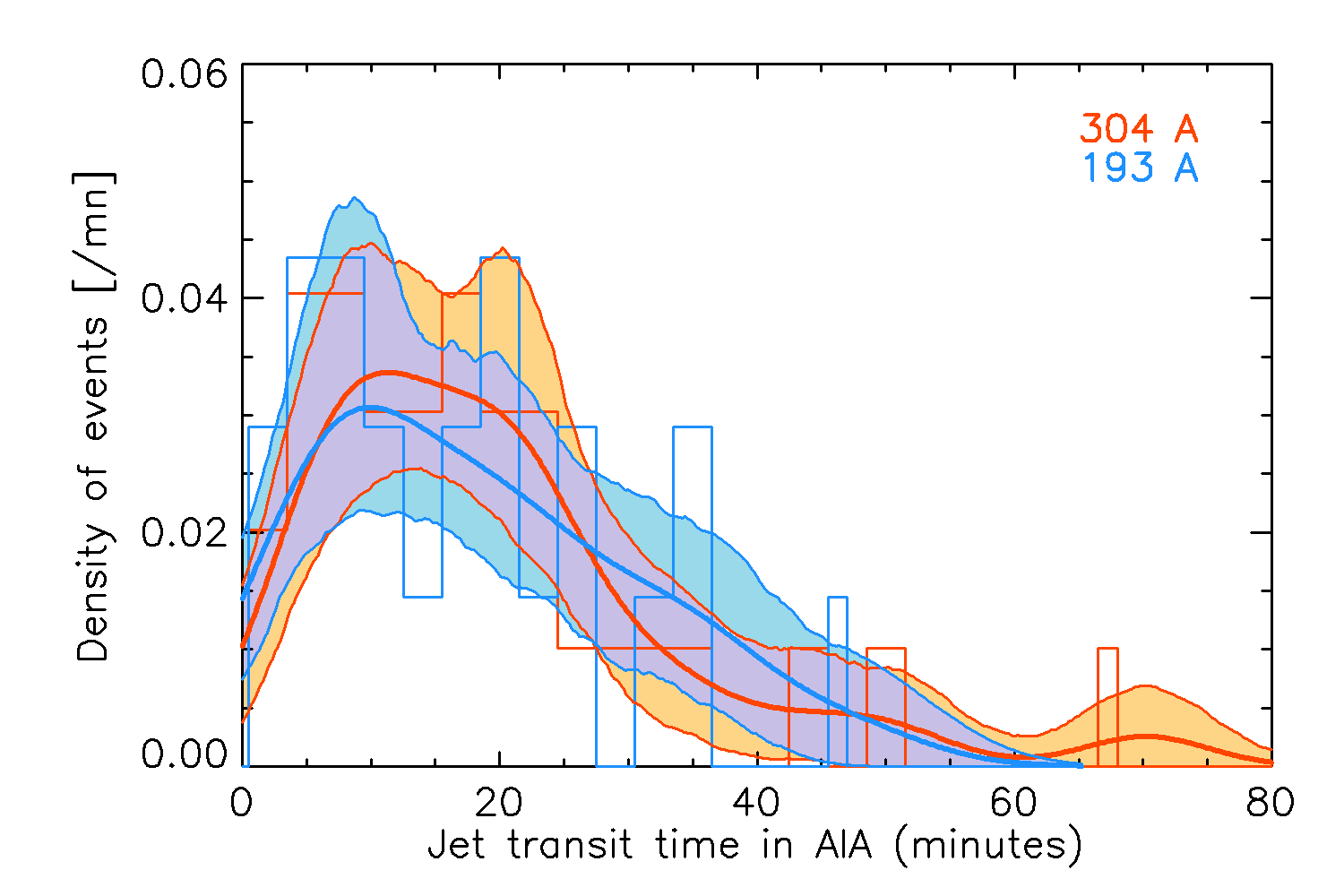}
\caption{Distribution jet durations (transit time in the region selected) in minutes, calculated from AIA 304 \AA \ (in red) and 193 \AA \ data (in blue). Histograms are shown as steps of 3 minutes. The kernel density estimations of the distributions are shown in thick lines. The 95 \% confidence intervals on this density estimation are shown as shadowed regions. }
\label{jet_duration_histo}
\end{figure}

\begin{table}[t]
\centering
\begin{tabular}{cccccc}
\hline
 & \textbf{Mean} & \textbf{Std. dev.} & \textbf{Median} & \textbf{Min.} & \textbf{Max.} \\ \hline
$D_{jet}^{304}$   &  20.2   &  14.3   &  18.8   &  4.2   &  70.2    \\ 
$D_{jet}^{193}$   &  18.6   &  12.1   &  18.8   &  2.8   &  47.2     \\ \hline
\end{tabular}
\caption{Statistics of the jet durations (in minutes) $D_{jet}^{304}$ and $D_{jet}^{193}$  calculated with the AIA 304 and 193 \AA \ respectively: mean value, standard deviation, median, minimum and maximum values. }
\label{table:durations}
\end{table}

The distribution of jet durations obtained for our sample is shown in figure \ref{jet_duration_histo}. Durations were estimated in both the 304 \AA \ and the 193 \AA \ data sets. The kernel density estimation are calculated using the R \citep{R_ref} generic function \textit{density} with a Gaussian kernel and the banwidth calculated with Silverman's rule of thumb \citep{silverman_1986}. 
The 95 \% confidence intervals are calculated with a bootstrap method. 
The statistics of the distributions is described in table \ref{table:durations}. The two distributions are very similar to each other, and exhibit a median jet duration of 18.8 minutes. The distribution seems to fall down towards zero minutes, which is due to the fact that given the 12-second cadence of AIA, we do not expect to detect events with short durations, practically, we do not expect to detect events shorter than one minute.
The calculated durations are reported in the list in appendix \ref{sec:listjets}.

This automatic detection of the jet start and end times leads to some uncertainty on the timing, due to the particular choice of the region of interest, which sometimes does not include the base of the jet, to avoid contamination of the jet lightcurve by the flare emission. We therefore checked for a bias in our start time determination, by also reporting the start time of each jet determined by visual inspection of the jet movies. We found that the start time of the jet could differ by one or two minutes on average when using these different methods. However, we did not find any systematic tendency for the automatically calculated start time to be later than the start time reported by eye, and the results regarding delays between jet and flare start times were not significantly changed when using start time reported by eye. We conclude that this automatic detection does not introduce a significant bias or error on the value of the start time of the jet, in comparison with a report of the time by a human looking at a movie of the jet.

\subsubsection{Jet velocity}
\label{sub:jetvelo}

\begin{figure*}
\includegraphics[width=\linewidth]{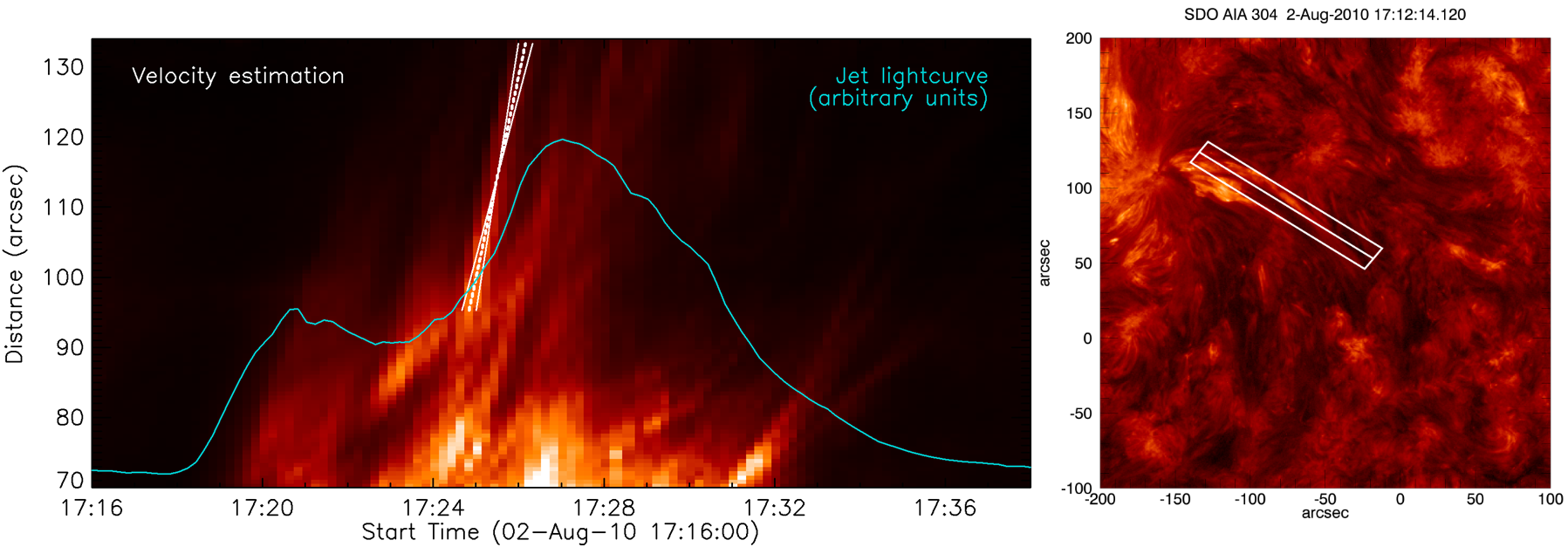}
\caption{Time-distance plot for the 2010 August 02 jet. The insert show the region and line used to create this plot. The white lines on the time-distance plot represent the two limits for the slope used to calculate the velocity of the jet. }
\label{time_distance}
\end{figure*}

\begin{figure}
\includegraphics[width=\linewidth]{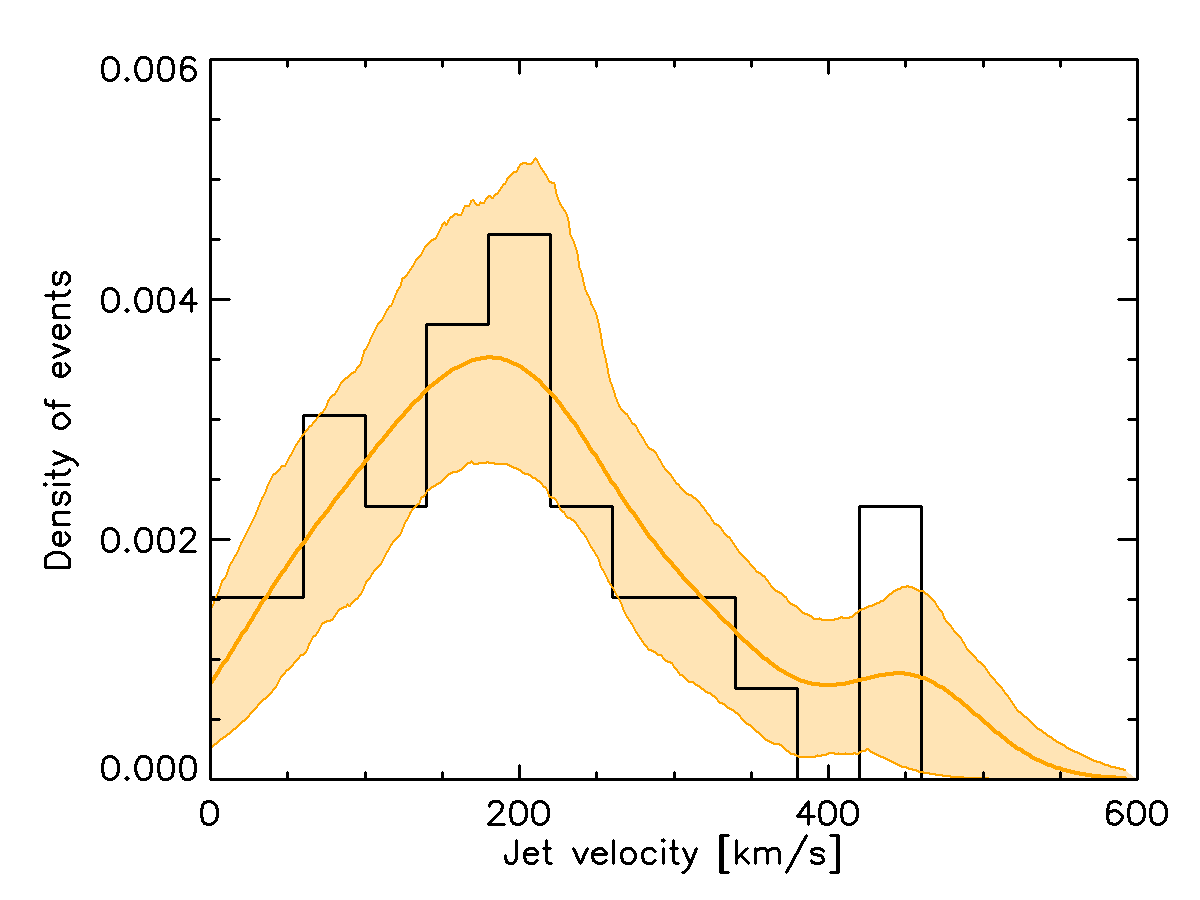}
\caption{Distribution of the jet velocities calculated in the 304 \AA \ AIA channel. The orange curve show the kernel density estimation with a 95\% confidence interval shown in pale orange. }
\label{velocity_distribution}
\end{figure}

\begin{table}[t]
\centering
\begin{tabular}{cccccc}
\hline
              & \textbf{Mean} & \textbf{Std. dev.} & \textbf{Median} & \textbf{Min.} & \textbf{Max.} \\ \hline
$V$           & 207    & 116    & 210    & 31    & 456       \\ \hline
$\delta V$    &   0.22 &   0.18 &   0.16 &  0.04 &   0.76    \\ \hline
\end{tabular}
\caption{Statistics of the jet velocities $V$ (in km/s) and relative errors ($\delta V = E_V / V$ where $E_V$ is the measured error) calculated with the AIA 304 \AA \ channel: mean value, standard deviation, median, minimum and maximum values. }
\label{table_velo}
\end{table}

For each of the jets, their velocity in the plane of the sky was calculated with time-distance plots computed with the AIA 304 \AA \ observations. A narrow rectangular region was selected around the jet, as shown in the right panel of figure \ref{time_distance}. For each image, the pixels are summed in the width of that region to obtain a one-dimension vector containing the intensity of the jet along its direction of propagation. This measurement is approximating the intensity of the jet as a function of the distance from the Sun surface. An example of the time-distance plot obtained is display in the left panel of figure \ref{time_distance}. Such plot is used to calculate the height evolution over time, and then the jet velocity. It must be noted that the velocity estimated from this technique is affected by projection effects, that are impossible to correct with observations in only one line of sight. Therefore, the velocity calculated here is a lower limit on the true velocity of the plasma ejection.
As it can be seen in figure \ref{time_distance}, an event defined as an individual jet in this study can be composed of several pulses of ejection, which are difficult to disentangle from one another.
The velocity of the jet is calculated by selecting by hand the beginning and end points of a pulse in the time-distance plot (see thick dotted white line in the left panel of figure \ref{time_distance}). For each jet, we selected the higher of the bright pulses. The error on the velocity is estimated to be of one pixel in the time direction, which is $\pm 12$ seconds (see plain thin lines in figure \ref{time_distance}). 

The distribution of jet velocities is shown in figure \ref{velocity_distribution}. The statistics of the distribution is reported in table \ref{table_velo}. The mean value of the velocities is 207 km/s, and the average relative error is 22 \%. 

\subsection{Jet relation with RHESSI flares}

The Reuven Ramaty High Energy Solar Spectrometer and Imager \citep[RHESSI,][]{rhessi} was a solar spectro-imager equipped with 9 germanium detectors and 9 associated rotating collimators. The imaging principle based on signal modulation is an indirect method for X-ray imaging. RHESSI is able to provide imaging of flares with a spatial resolution of 3 arcsec, and spectral information from 4 keV to MeV range, allowing to probe both thermal and non-thermal X-ray emission in solar flares. The mission was launched in 2002 and decommissioned in 2018 and therefore covers the SDO lifetime until recently.

\subsubsection{GOES flux of flares}

Not all of our events are present in the GOES flare list, probably because they are associated \textbf{with} small flares which size is at the detection limit of GOES. For each jet in the list, we calculated the background-subtracted GOES flux and estimated the peak flux of the jet-associated flare.
We used the GOES-15 lightcurve at 1-8 \AA \ for the whole sample.

For each jet-associated flare, the subtracted background GOES flux was computed semi-automatically:
\begin{enumerate}
\itemsep0em 
\item The time interval considered is centered on the flare peak time and extends $15$ minutes before and after the flare as reported in the RHESSI flare list. 
\item The background is evaluated as the intervals for which the derivative of the GOES lightcurve is close to $0$ and the value of the lightcurve does not exceed the linear regression of the lightcurve.
\item A linear regression is performed over those intervals to estimate the background.
\item The estimated background is removed from the lightcurve.
\item The GOES peak is found in the flare interval and the peak flux is reported for the background-subtracted data.
\end{enumerate}
This process is only semi-automated: if the result background evaluation was not done properly on visual inspection, a different time interval was selected until the result was acceptable when checking the curves. Note that in that process, variable background is assumed to be linear, which is not always the case, but is a valid approximation in the scope of this paper.

\begin{figure}
\includegraphics[width=\linewidth]{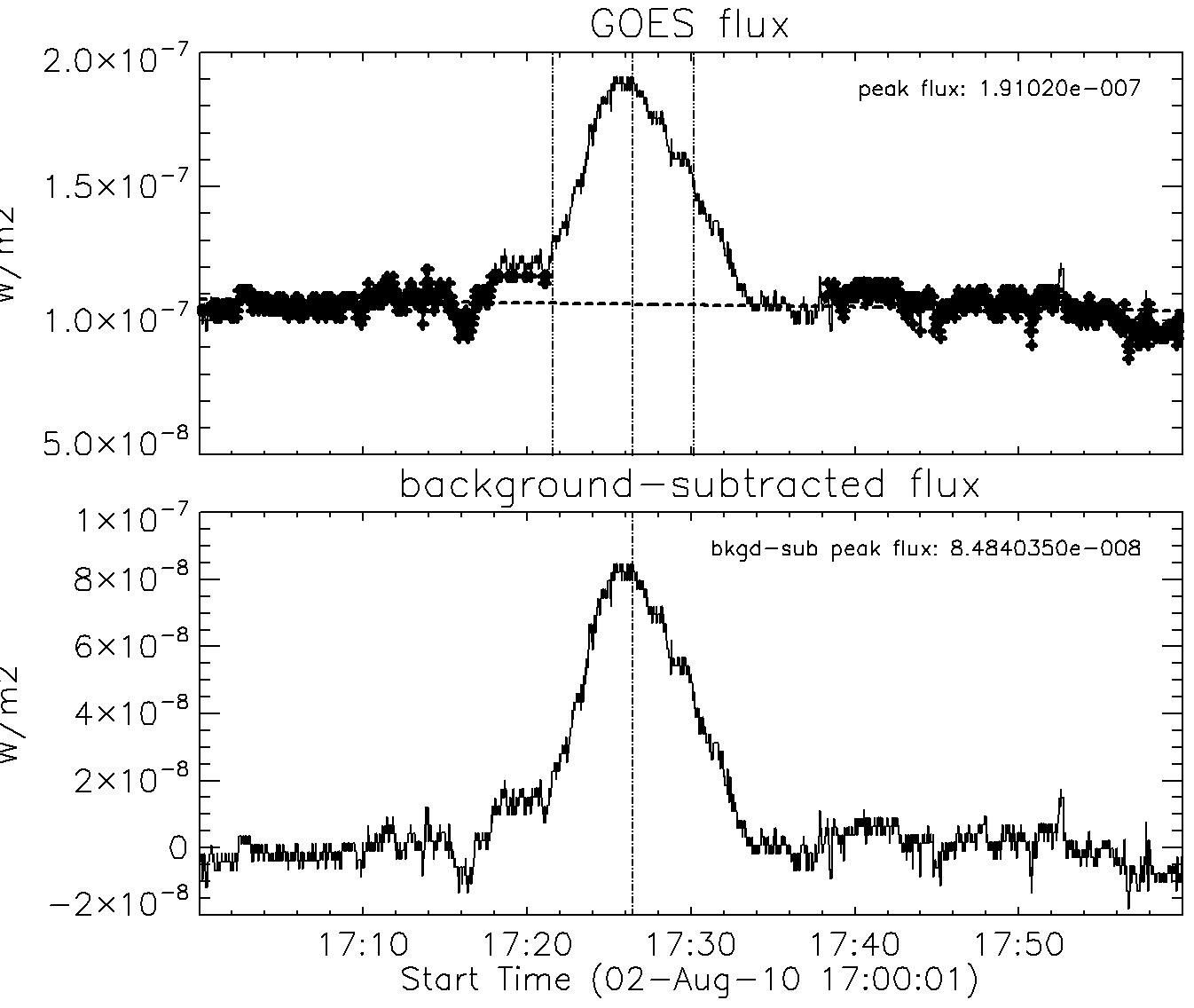}
\caption{Top: GOES lightcurve at 1-8 \AA . The thick line is where the background was evaluated, the horizontal dashed line is the background estimation. The vertical lines represent the beginning of the RHESSI flare, the GOES peak, the end of the RHESSI flare. Bottom: background-subtracted GOES flux. The horizontal line is at the GOES peak time.}
\label{goes_class}
\end{figure}

An example of this process is illustrated in figure \ref{goes_class} for the 2010 Aug 02 jet.

\begin{figure}
\includegraphics[width=\linewidth]{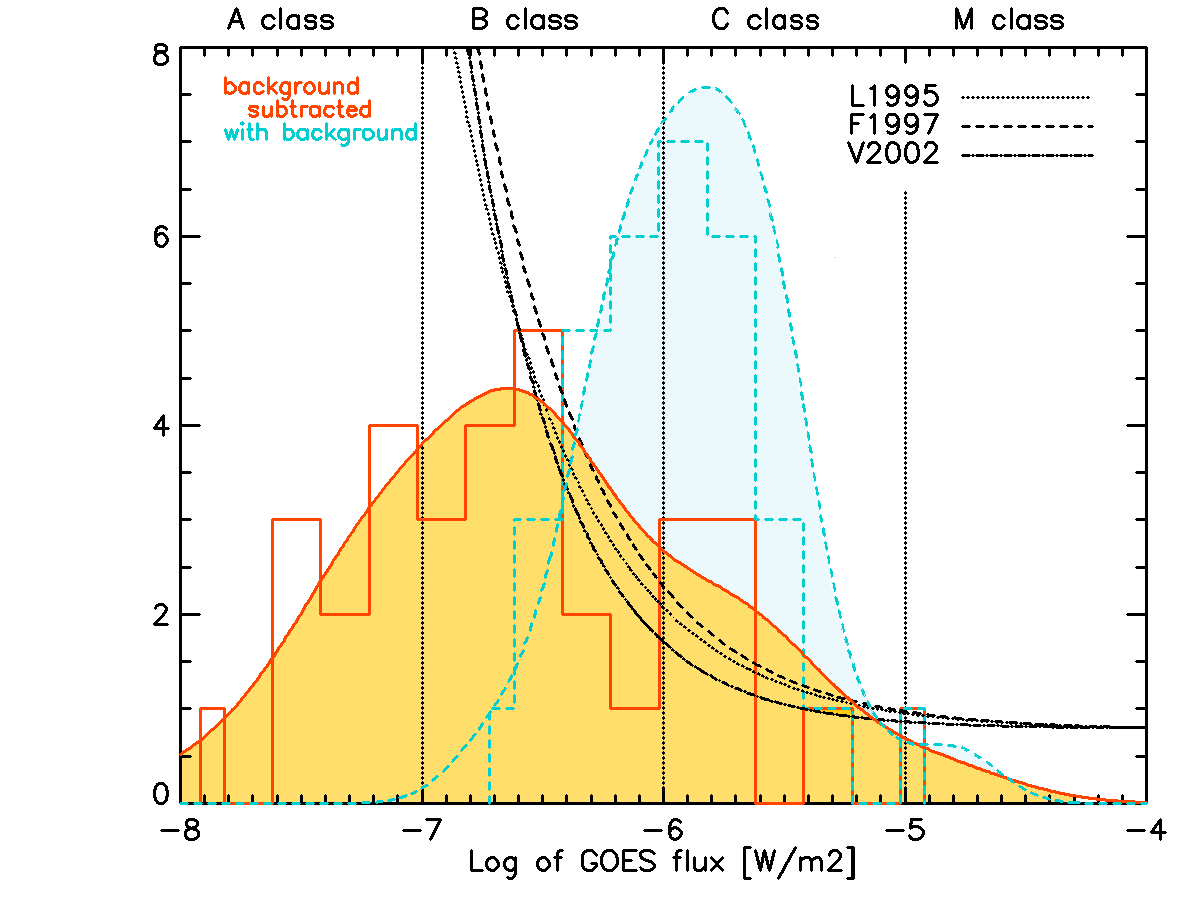}
\caption{Distribution of GOES peak fluxes in 1-8 \AA , for jet-related flares. The orange plain histogram show the distribution of background-subtracted GOES peak fluxes, while the dashed blue histogram show the distribution of GOES peak fluxes without the background subtraction. The plain orange and dotted blue curves show the kernel density estimation of the distributions. The black dashed-dotted curves show the trends of the distributions computed in \cite{lee_etal_1995} (L1995), \cite{feldman_etal_1997} (F1997), and \cite{veronig_etal_2002} (V2002). The dotted vertical lines show the separation between B-class, C-class and M-class flares.}
\label{goes_class_histo}
\end{figure}

The distribution of GOES flux obtained for our list of event is given in figure \ref{goes_class_histo}. Most of the flares are GOES B class. The distribution is falling down towards small flares (A size) most likely because of the sensitivity of instrument. 

\subsubsection{Relative timing and flare and jets}

The delays between jets and flares can be examined in different ways. 
First, one can examine the delays between the start and peak times of the jet and of the flare. Secondly, a cross-correlation of the jet and flare lightcurve can be performed in order to identify the delay of the jet in regards to the flare.

\begin{figure}
\includegraphics[width=\linewidth]{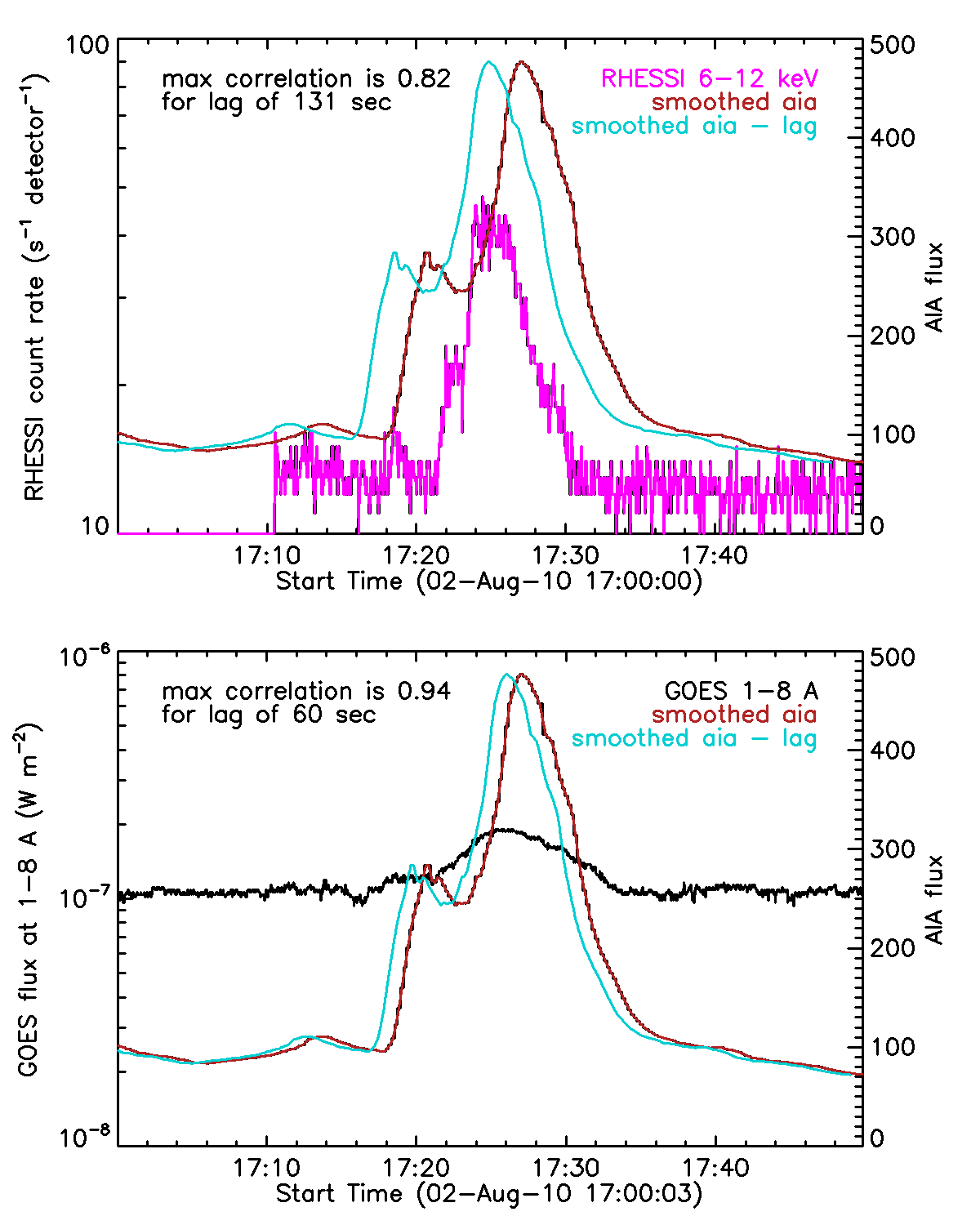}
\caption{Cross-correlation between the AIA 304 \AA \ lightcurve of the 2010 Aug 02 jet with the RHESSI 6-12 keV lightcurve (top) and the GOES 1-8 \AA \ lightcurve (bottom). The best correlation is obtained for a delay of 131 seconds between the flare and the jet, with the RHESSI ligthcurve, and  a delay of 60 seconds between the flare and the jet, with the GOES lightcurve.}
\label{cross_correlation}
\end{figure}

The flare peak time is the time of maximum intensity in a given energy range. In this paper we considered the peak time in the GOES lightcurve at 1-8 \AA \ and the peak time in the RHESSI 6-12 keV energy band. Both these energy ranges reflect the thermal behavior of the flare. When a non-thermal component was detected (see section \ref{sec:spectro}), we also considered the flare peak time in 12-25 keV or 25-50 keV, depending on the energy at which the transition between thermal and non-thermal emission was found.
The flare start time was taken from the RHESSI flare list and therefore has been automatically generated. It corresponds to the time of detection of a rise of the 6-12 keV emission above the background level. We note that given the high background in RHESSI detectors, this RHESSI start time of the flare may be later that the real start of the flare.

The jet start time has been automatically calculated as described in section \ref{sec:jets_timing} using the jet lightcurve computed in EUV. The jet peak time corresponds to the time of maximum intensity in the jet lightcurve.

When the jet lightcurves and flare lightcurves have similar shapes, a cross-correlation is performed in order to identify the time delay needed to maximize the correlation between the curves: an example is shown in figure \ref{cross_correlation}. The result of this cross-correlation was kept only when the cross-correlation was converging with a correlation coefficient greater than 0.50. This was successfully performed for 20 events on the list.

Overall, we computed the delay between the peak of the flare and the peak of the jet in both 304 \AA \ and 193 \AA \ wavelengths, as well as the delay between the start of the flare and the start of the jet. For the flares for which we detected non-thermal emission (see section \ref{sec:spectro}), we also looked at the delay between the peak of the non-thermal flare emission and the start and peak times of the jet. The results of those are described in section \ref{discussion:timing}.


\subsubsection{X-ray imaging}
\label{sec:imaging}

\begin{figure*}
\includegraphics[width=\linewidth]{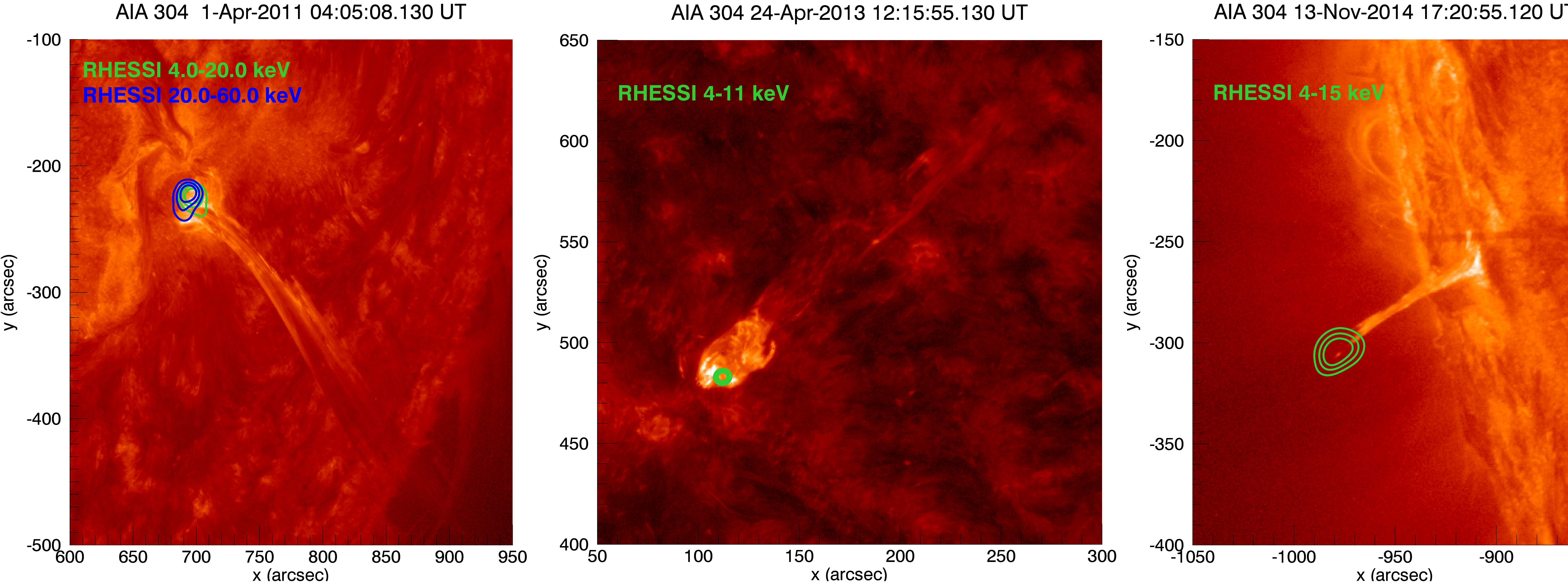}
\caption{Example of RHESSI emission in flare-related jets. Left: the RHESSI emission is at the base of the jet, and non-thermal emission is detected. Middle: Thermal X-ray emission at the base of the jet. Right: unusual situation, where the X-ray thermal emission is emitted at the top of the jet. The jet images are from AIA 304 \AA \ channel.}
\label{images}
\end{figure*}

RHESSI imaging was performed for each flare around the flare peak time. The integration time, the energy range and the choice of collimator were adjusted for each flare. We used the CLEAN and the Visibility Forward Fit algorithms. The Visibility Forward Fit is performed by fitting parameters corresponding to a source shaped as a curved gaussian (a loop) to the RHESSI visibilities. This method provides a source length and width with an error estimation and is therefore used to estimate the volume of the RHESSI source. The images produced with the Visibility Forward Fit have been visually compared to the CLEAN images to ensure the consistency between the two imaging algorithms. A more quantitative comparison between the volumes calculated with the CLEAN images and with the Visibility Forward Fit is discussed in appendix \ref{sec:volumes}. The images are made in the thermal energy range of each flare and therefore provide the location and the size of the source of thermal X-ray emission. 
The thermal emission of the jet is not particularly expected to match the location of the base of the jet, as it could in principle be emitted by a side magnetic loop reconnecting with the open magnetic field lines carrying the jet. However, it is also possible that the thermal loops are too small to be clearly resolved and appeared to be at or close to the base of the jet.
The thermal volume $V$ of the X-ray source assuming a cylindrical geometry, and a filling factor equal to 1, is calculated using:
\begin{equation}
V =  \pi \left( \frac{w}{2} \right)^2 l
\end{equation}
where $w$ is the width at the middle of the loop, and $l$ is the FWHM loop arc length.

Examples of the RHESSI images superimposed on the jet images at 304 \AA \ are shown in figure \ref{images}, and the images for the remaining jets are shown in appendix \ref{sec:alljets}. In general X-ray sources are located at the base of the jet, with sometimes some extension in the jet direction, as shown on the left and middle panels. In more rare cases, X-ray emission comes from the jet itself (see for example right panel of figure \ref{images}). 
In the studied cases, only thermal emission has been identified to be emitted from the jet itself. When non-thermal emission is observed, it is generally located at the base of the jet or in an adjacent loop (mostly at the footpoints).

\subsubsection{X-ray spectroscopy}
\label{sec:spectro}

We performed HXR spectroscopy on each of the RHESSI flares associated with the jets. For most events, we chose the spectroscopy time interval to encompass the entire event between the flare start and end times listed in the RHESSI flare list.  We also inspected light curves for each event and, in some cases, chose a time interval centered on a high-energy peak in the light curve (if one was present) or just after an attenuator insertion to avoid pile-up effects.  
Background intervals were chosen to be 4-minute intervals during the nearest eclipse to the flare.  In cases where data was not recorded during eclipses, or when SAA and eclipse intervals overlapped, background intervals were chosen by eye just before or after the flare.  
Spectrograms for every detector were examined in order to choose the most appropriate set of detectors in each case; detectors that exhibited obvious issues were excluded.

Two models were fitted to each flare spectrum: (1) an isothermal model (termed vth in the RHESSI OSPEX spectral fitting software); and (2) an isothermal component plus a broken power-law (vth+bpow). 
Case (2) allows for the presence of detectable non-thermal electrons.  
Standard coronal abundances were used.  
The spectral index of the power-law component below the break energy was fixed at 1.7 as this always lay below the thermal component and could not be well fit.  
The lower energy threshold for fitting was 4 keV, unless attenuators were inserted (in which case a higher threshold of either 6 or 10 keV was used).  
The fit energy range extended up to the energy at which the spectrum fell below 10 counts per energy bin.

\begin{figure}
\includegraphics[width=\linewidth]{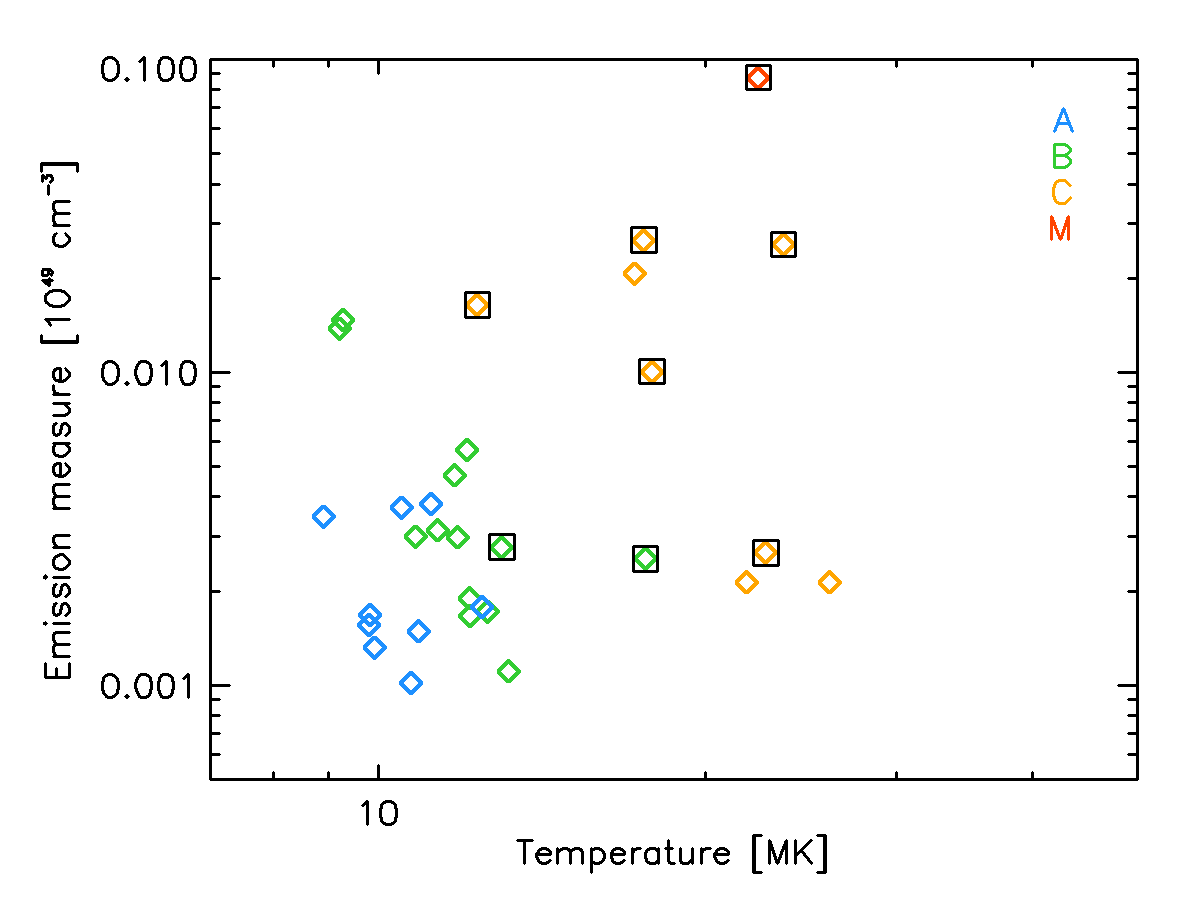}
\caption{Distribution of the flare temperature and emission measure. Colors reflect the GOES class calculated with background-subtracted GOES peak flux, and the flares for which a non-thermal component has been detected are squared in black.}
\label{spectro_thermal}
\end{figure}

Events presenting a non-thermal component were detected by looking at the thermal fits to the count spectra. They correspond to events for which the thermal fit gave a reduced chi-squared superior to 40, except for one outlier that had a chi-squared of 33 but still exhibit a non-thermal tail in the spectrum when examined. 
When a non-thermal component is present, the slope index $\delta$ is reported in the jet list (table \ref{listofjets} in appendix \ref{sec:listjets}). We found non-thermal emission for 8 events over 33, which is 1/4 of the events. The values of the photon spectral indexes vary from -1.8 to -4.3. The results of the thermal fits give the temperature and emission measure of the X-ray emitting plasma. Those results are summarized in figure \ref{spectro_thermal}.


\section{Statistical analysis} \label{sec:analysis}

\subsection{What is the size distribution of jet-related flares?}

The idea that jets could be seen as small-scale filament eruptions or coronal mass ejections suggests that such events should be associated with smaller flares. 
Indeed, if size and velocity scale between jets and CMEs, their potential and kinetic energy will scale as well, jets being at the low-end of their distribution. One can imagine in this case that the energy of the associated flare also scales, in particular their radiative energy.
We used the peak GOES flux at 1-8 \AA \ as the measurement of flare size. The resulting distribution of flare sizes is shown in figure \ref{goes_class_histo}.

The distribution of the number of solar flares at different peak intensities in the 1-8 \AA \ GOES range has been examined by several studies. The frequency distribution can be represented by a powerlaw of the form
\begin{equation}
{dN} = {A F^{- \alpha}} dF  
\end{equation}
where $dN$ is the number of flares recorded with a GOES peak flux between $F$ and $F + dF$, $A$ and $\alpha$ are constants to be adjusted when fitting the flare distribution. When integrating this expression, we obtain the expected number of flare in each bin of the histogram displayed in figure \ref{goes_class_histo}:
\begin{equation}
N_i = A' F^{-(\alpha-1)}
\end{equation}
where $A'$ is a constant related to $A$, $\alpha$ and $dF$.

Using this expression, we are able to trace the distributions $N_i$ (with arbitrary values of constants $A'$) for the values of $\alpha$ found in \cite{lee_etal_1995, feldman_etal_1997, veronig_etal_2002}: $1.86 \pm 0.10$, $1.88 \pm 0.21$ and $2.11 \pm 0.13$ respectively. Note that the value found by \cite{veronig_etal_2002} has been computed on a distribution of the GOES peak values for which the background was not subtracted, and is therefore expected to be greater than values obtained for background-subtracted values, since background subtraction flatten to steepen the distribution. Those distribution are visible in figure \ref{goes_class_histo} (black lines). 
To quantify how these distributions agree with our own data set, we performed a Kolmogorov-Smirnov test including in the GOES flux range $10^{-6.8} - 10^{-3} W/m^2$, since our distribution fall down to smaller fluxes, due to the limit of the sensitivity of the GOES instrument, 
and we found that our distribution is in agreement with the previous results, with a probability of the order of 60 \%.

\subsection{Is there a link between flare size and jet properties?}
\label{discussion:flaresize}

\begin{figure}
\includegraphics[width=\linewidth]{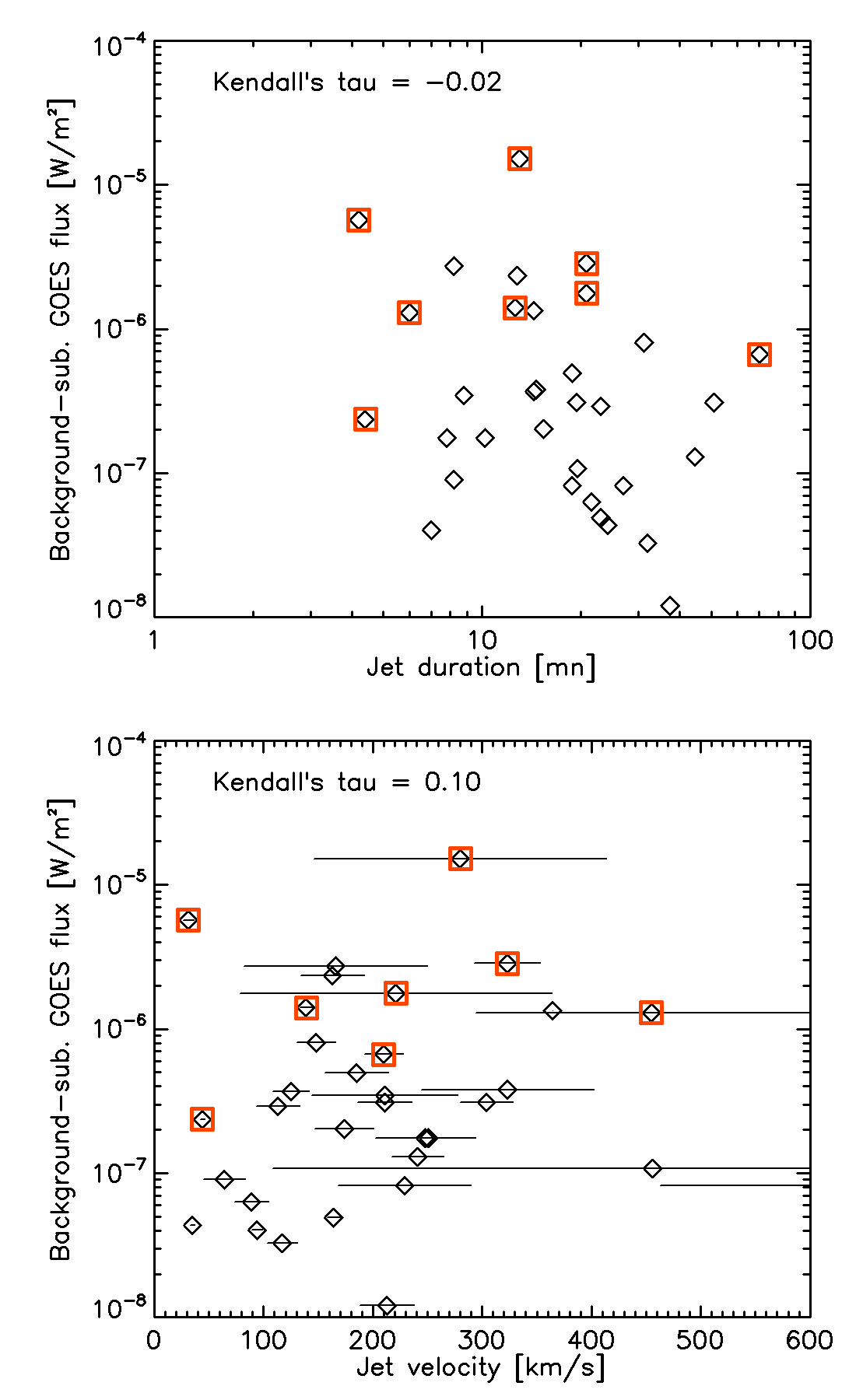}
\caption{Top: Distribution of GOES peak fluxes in 1-8 \AA , for jet-related flares, in relation with the jet duration, as seen in 304 \AA \ images. Bottom: Distribution of GOES peak fluxes in 1-8 \AA , for jet-related flares, in relation with the jet velocity projected in the plane of sky, as seen in 304 \AA \ images. The horizontal bars show the error on the velocity measurements. In both panels, the events for which we observed non-thermal X-rays are shown in red squared symbols. For both data set, the nonparametric Kendall's tau coefficient is given.}
\label{jet-goesclass}
\end{figure}

We investigate the possibility of a link between the energy of jets and the energy of the associated flare. 
In this study, we look at the flare size (background-subtracted GOES peak flux) and its thermal energy. 
Top panel in figure \ref{jet-goesclass} shows the relation between the flare size and the jet duration. 

\textbf{To establish a possible correlation between two independent variables, we used the nonparametric Kendall's tau coefficient, which tests the correlation without any assumption on the form of the correlation, and is based on the comparison of ranks in the two sets of variables.}
The nonparametric Kendall's tau coefficient for these data is -0.02, \textbf{and associated to a p-value of 0.87}, demonstrating that there is no correlation between the flare size and the jet duration. 
Bottom panel in figure \ref{jet-goesclass} show that there is no correlation between the flare GOES class and the jet velocity: the Kendall's tau  is 0.10 for these data.

\begin{figure}
\includegraphics[width=\linewidth]{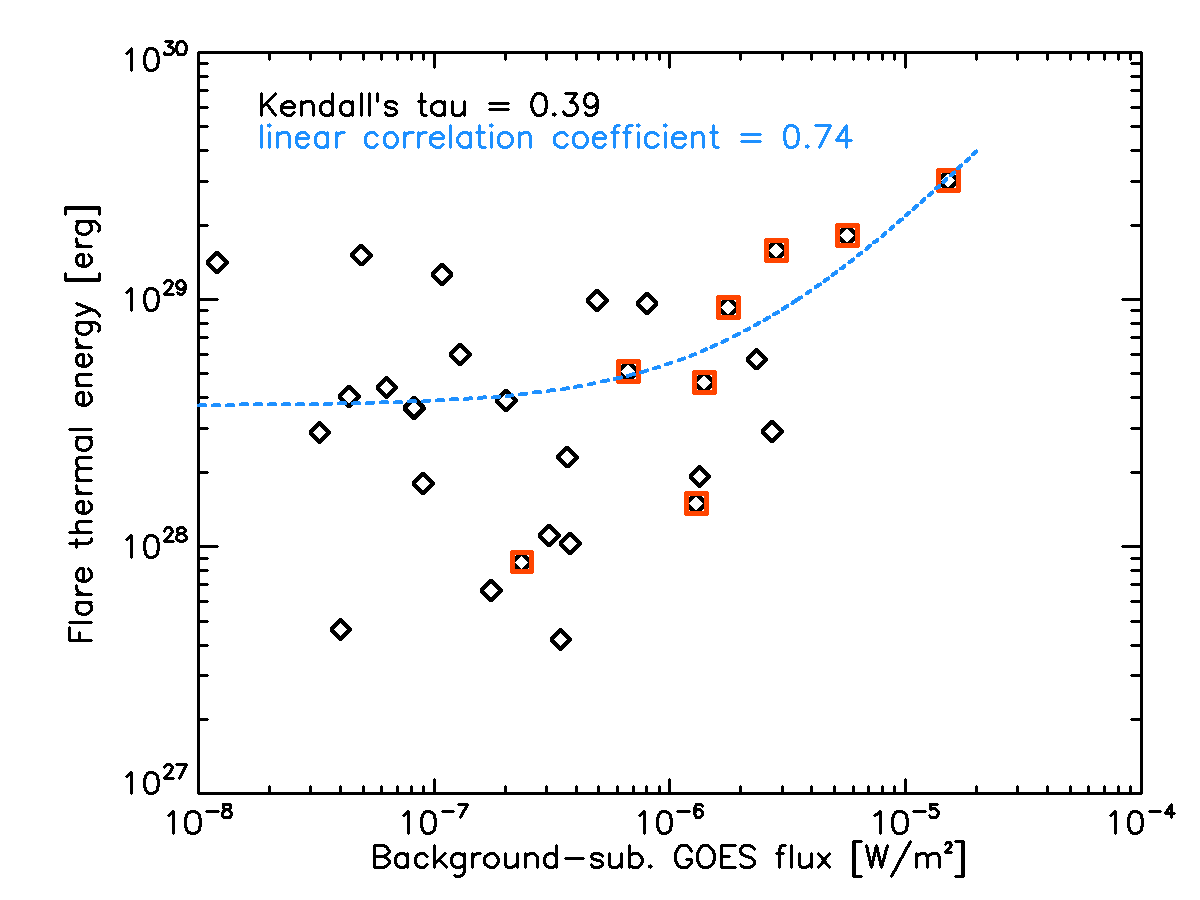}
\caption{Distribution of the flare thermal velocity deduced from X-ray spectroscopy, in relation with the GOES subtracted peak flux (GOES class) of the flare. The result of the linear correlation between the flare thermal energy and the background-subtracted GOES flux is shown as a dotted blue line. The red square data points show the events for which non-thermal emission is detected.}
\label{goes_thermal}
\end{figure}

The thermal energy of the flare was estimated using the result of the isothermal fit to the X-ray thermal spectrum. The parameters of the thermal component of the fit are the plasma temperature $T$ and emission measure $EM$.
Those values can be used to calculate the thermal energy of the flare, using the relation:

\begin{equation}
W_{th} = 3 \sqrt{EM V} k_B T
\end{equation}
where $V$ is the source volume determined from the X-ray images, described in \ref{sec:imaging}.

\begin{figure}
\includegraphics[width=\linewidth]{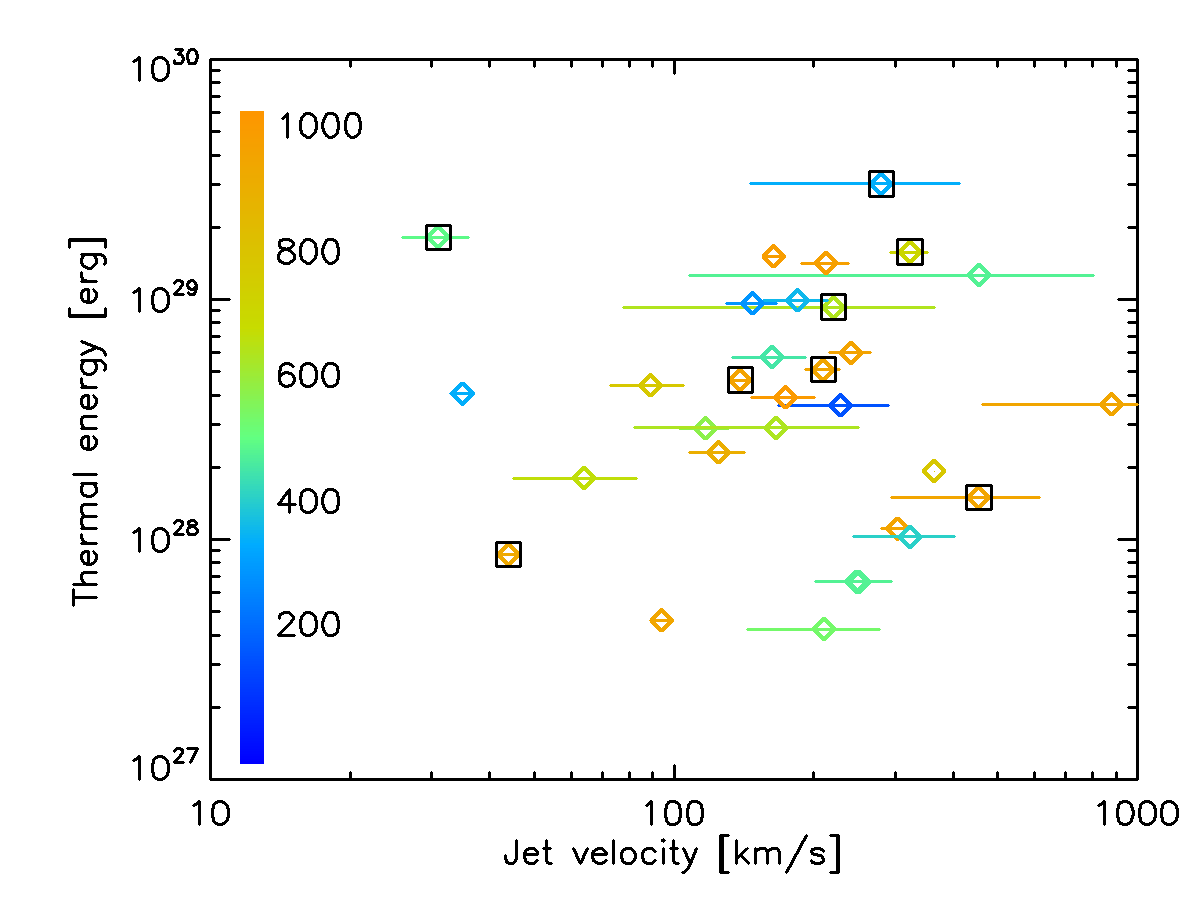}
\caption{Distribution of the flare thermal velocity deduced from X-ray spectroscopy, in relation with the jet velocity projected in the plane of sky, as seen in 304 \AA \ images. The color scale represents the radial distance of the base of the jet from disk center, is arcseconds (color bar). The horizontal bars represent the measurement error on the value of the projected velocity. The non-thermal flares are enlightened with a black square.}
\label{velocity_thermal}
\end{figure}

As shown in figure \ref{goes_thermal}, there is a good correlation between the thermal energy of the flare deduced from RHESSI observation and the GOES class of the flare. The nonparametric Kendall's tau value is 0.39, with an associated p-value of 0.003, which show a relationship between the two quantities at the 0.01 significance level. Assuming a linear correlation between the thermal energy and the GOES peak flux, the correlation coefficient is 0.74. 
The distribution of GOES background-subtracted fluxes scatter at the lower end of the distribution. This is due to the fact that the GOES signal becomes less accurate below the B7 level, as shown with simultaneous measurements of the soft X-ray flux with the MinXSS instrument \cite[see figure 4 in][]{woods_etal_2017}.  
The flares for which non-thermal X-ray radiation has been detected are shown in red on figure \ref{goes_thermal}: they lies in the end of the distribution of GOES fluxes. However, the thermal energy of these events seems scattered. It must be noted that the four events with smaller thermal energy are on the limb of the Sun. It is therefore possible that the volume of the source, and therefore the thermal energy, is underestimated for those events. However, as shown in figure \ref{velocity_thermal}, there is no obvious bias towards smaller thermal energy estimates for near-the-limb events (shown in orange color).

 The Kendall's tau coefficient for the relation between the jet velocity and the flare thermal energy (figure \ref{velocity_thermal}) is 0.04, and is associated with a p-value of 0.73, which shows that the thermal energy and the kinetic energy of the jet are not correlated.

\subsection{What is the relative timing between flare and jets?}
\label{discussion:timing}

\begin{figure}
\includegraphics[width=\linewidth]{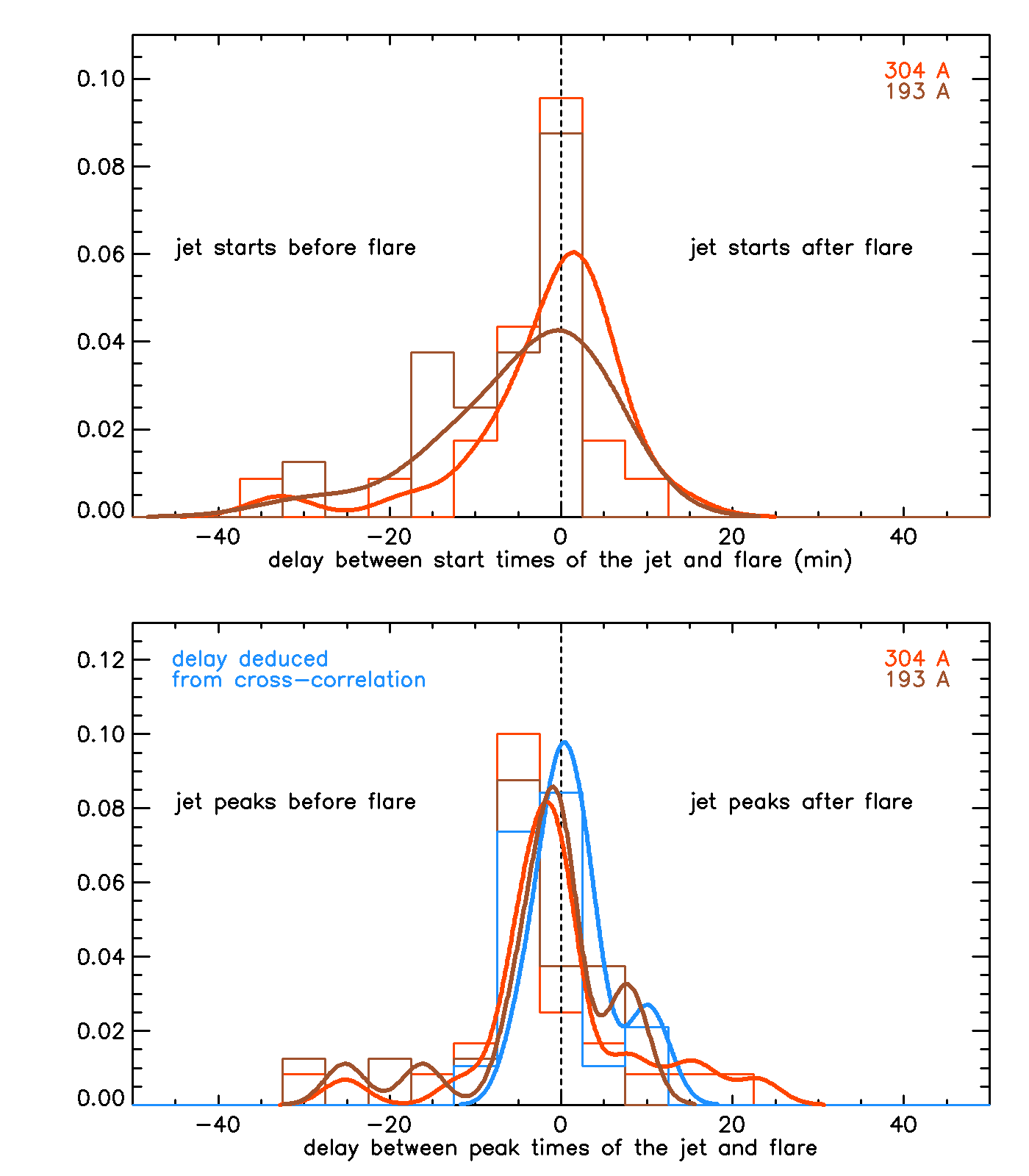}
\caption{Top: delays between the jet and the flare start times. Bottom: delays between the jet and flare peaks times and delays deduced from the cross-correlation between the jet 304 \AA \ lightcurve and the 6-12 keV RHESSI lightcurve. Thick lines represent the kernel density estimation deduced with the R \textsl{density} function with a Gaussian kernel and a bandwith of 1.5 times the bandwith deduced with Silverman's rule of thumb.}
\label{jet_delays}
\end{figure}

The delays between the flare and the jet have been calculated with different methods and different characteristic times, as explained in section \ref{sec:jets_timing}.
The first interesting result that can be seen on figure \ref{jet_delays} is that the distribution of delays do not vary significantly when we look at the jet in 304 \AA \ (colder plasma emission, with characteristic temperature of $10^{4.7}$ K) or in 193 \AA \ (hotter plasma emission, with characteristic temperatures of $10^{6.2}$ and $10^{7.3}$ K). The top panel in figure \ref{jet_delays} shows that the jet and flare mostly start at the same time, with a tail of the distribution showing a slight tendency to see the jet starts before the flare. 
However, it must be noted that RHESSI is not a low-background instrument. Therefore, the reported start time of a flare can be seen as an upper limit, the flare could be starting earlier than reported, with X-ray emission not yet detectable. This could explain a slight delay of the flare, probably of the order of a few seconds to a few minutes.

We also note that a slight shift between the peak of the distributions of the delays deduced from peak times and the peak of the distribution of delays deduced from a cross-correlation can be seen, in the bottom panel of figure \ref{jet_delays}. However, the
12-seconds cadence for the EUV data of the jet and the small number of events in our study limit our capacity to draw more precise conclusion than the fact that if there are delays between the flare and the jets, they are in general smaller than 2 minutes, and there is no clear tendency for the jet to precede or follow the flare.

\begin{figure}
\includegraphics[width=\linewidth]{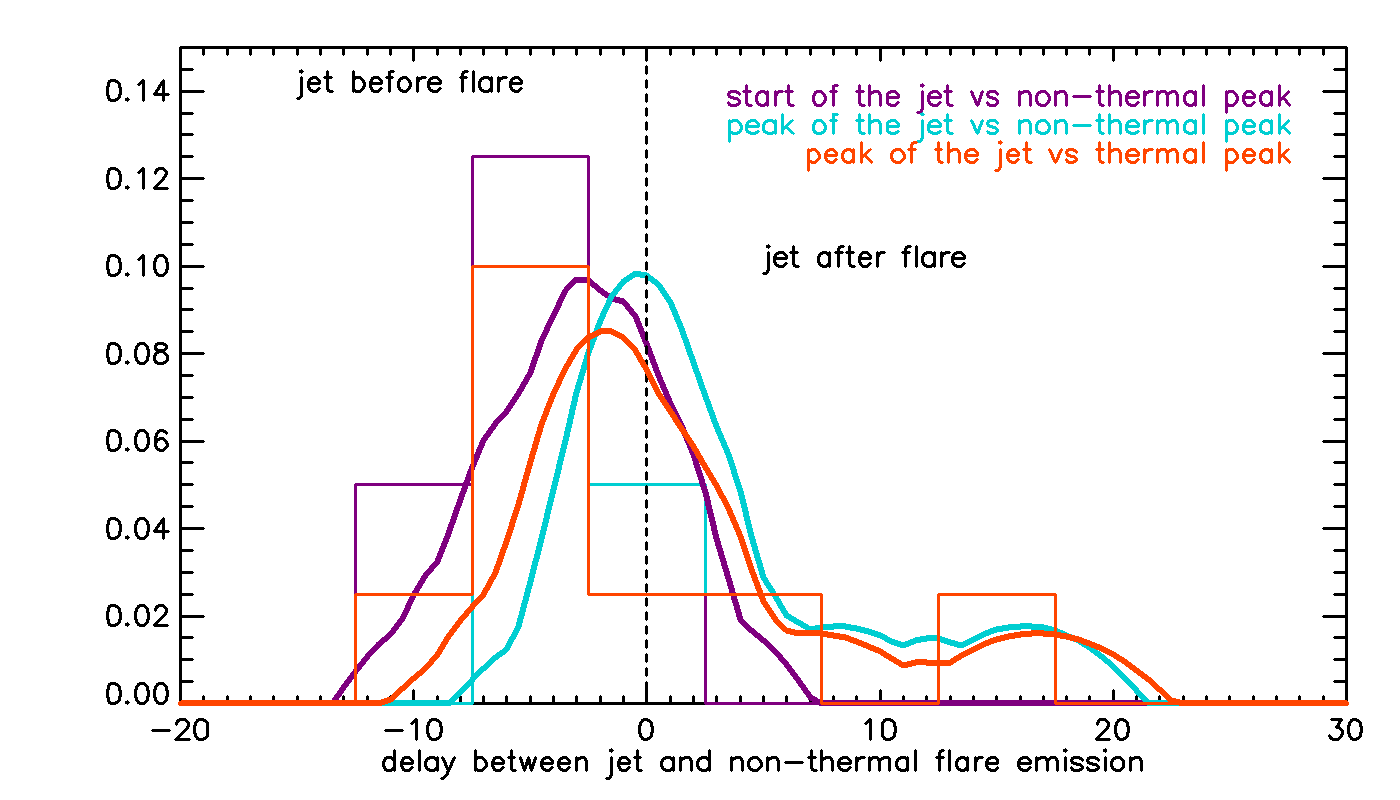}
\caption{Delay between the peak time of the non-thermal X-ray emission of the flare and the start time (in violet) or the peak time (in blue) of the jet, deduced from the 304 \AA \ emission. The delay between the thermal X-ray emission and the peak of the jet for the same sample of events in plotted in red.}
\label{jet_delays_nth}
\end{figure}

\begin{table}[t]
\centering
\begin{tabular}{cccccc}
\hline
\textbf{Description} & \textbf{$\mu$} & \textbf{$\sigma$} & \textbf{Med.} & \textbf{Min.} & \textbf{Max.} \\ \hline
$t_{s,jet}^{304}$ - $t_{s,flare}^{6-12}$      & -1.8 & 9.2  & 0.7    & -32.7   & 13.3    \\ \hline
$t_{s,jet}^{193}$ - $t_{s,flare}^{6-12}$      & -4.5 & 8.8  & -0.4   & -29.1   & 4.0     \\ \hline
$t_{s,jet}^{304}$ - $t_{p,flare}^{nth}$       & -4.0 & 3.2  & -4.4   & -8.2    & 1.4     \\ \hline
$t_{s,jet}^{193}$ - $t_{p,flare}^{nth}$       & -7.7 & 7.9  & -5.0   & -21.4   & -1.0    \\ \hline
$t_{p,jet}^{304}$ - $t_{p,flare}^{6-12}$      & 0.1  & 9.4  & -1.0   & -25.2   & 23.1    \\ \hline
$t_{p,jet}^{193}$ - $t_{p,flare}^{6-12}$      & -2.0 & 8.5  & -0.3   & -25.2   & 8.4     \\ \hline
$D_{cc}^{304}$                                & 1.7  & 4.5  & 0.2    & -5.1    & 11.6    \\ \hline
$t_{p,jet}^{304}$ - $t_{p,flare}^{nth}$       & 2.8  & 6.3  & 0.7   & -3.1    & 16.2    \\ \hline
$t_{p,jet}^{193}$ - $t_{p,flare}^{nth}$       & 3.1  & 4.4  & 1.0    & -0.9    & 8.1     \\ \hline
\end{tabular}
\caption{Statistics of the delays (in minutes) calculated between the flare and the jet: mean value ($\mu$), standard deviation ($\sigma$), median (Med.), minimum (Min.) and maximum (Max.) values. $t_{s,flare}^{6-12}$ and $t_{p,flare}^{6-12}$ are the start and peak times of the flare in the 6-12 keV energy range, respectively;  $t_{p,flare}^{nth}$ is the non-thermal peak of the flare; $t_{s,jet}^{304}$ and $t_{p,jet}^{304}$,  $t_{s,jet}^{193}$ and $t_{p,jet}^{193}$ are the start and peak times of the jet in 304 and 193 \AA \ respectively. $D_{cc}^{304}$ delay deduced from the cross-correlation between the flare lightcurve at 6-12 keV and the jet lightcurve at 304 \AA .}
\label{table:delays}
\end{table}
	
The little delay seen between jet and thermal emission of the jet can be seen as a confirmation that they both arise from the same episode of magnetic reconnection in the corona. For the subset of events for which we detected non-thermal emission, we used the peak of the non-thermal emission lightcurve as an additional characteristic time. As shown in figure \ref{jet_delays_nth}, there is a clear trend for the jet to start a few minutes before the peak of the non-thermal emission.
The peak of the jet emission will happen slightly before or at the same time as the peak of the non-thermal emission. 

\subsection{What fraction of jets show evidence of particle acceleration?} 

Non-thermal emission was detected in only 25 \% of the events. \textbf{The events with and without detected non-thermal emission are treated as two samples of data. We tested the null hypothesis of these two sample to be derived from the same population of flares using the Mood median test, which compares the medians of both samples.} Using the nonparametric Mood median test, we can conclude that non-thermal events correspond to the flares with the biggest peak GOES fluxes with a 5 \% significance level, but these events are not associated with the \textbf{biggest} values of of jet duration, jet velocities or flare thermal energies, as can be intuitively seen in figures \ref{jet-goesclass} and \ref{velocity_thermal}. Non-thermal emission is therefore associated with the largest flares, but not necessarily with the most energetic jets.


\section{Discussion} \label{sec:discussion}

\subsection{Jet velocity}

In the present study, we calculated the projected velocity of 33 coronal jets in the 304 \AA \ AIA channel. We found an average velocity of 207 km/s. This value is comparable with the mean value of 271 km/s found for 20 active region jets by \cite{mulay_etal_2016} and by the jet plasma velocity of 220 km/s calculated with spectroscopic observations of the He II emission line at 256 \AA \  by \cite{matsui_etal_2012} in one event. He II ions are the primary contributors to the AIA 304 \AA \ emission \cite{aia}, and the main plasma temperature contributing to this filter is $10^{4.9}$ K. 
The sound speed $c_s$ in a fully ionized plasma depends on the plasma temperature: $c_s = 147 \sqrt{T_{MK}}$ \citep[see e.g.][]{aschwanden_book}, and the sound speed for a  temperature of $10^{4.9}$ K is around 41 km/s. 
Therefore, our distribution of velocities is dominated by velocities higher that the sound speed for that plasma temperature. \cite{fisher_etal_1984} estimated that the upper limit of the speed for chromospheric evaporation was of $2.35 c_s$, which corresponds here to 97 km/s.
We note that only 6 jets in our study have a velocity smaller than 97 km/s.
We therefore come to the conclusion that the velocities measured are in average significantly higher that the sound speed and the predicted chromospheric evaporation speed, which is in agreement with the observations of \cite{matsui_etal_2012}. These observations seem to confirm that the He II emission outlines the propagation of Alfvenic perturbations (such) as MHD waves that drive the jet \citep{matsui_etal_2012, pariat_etal_2016}. 
The velocity deduced from the 193 \AA \ channel is not statistically different from the velocity deduced \textbf{from the} 304 \AA \ channel, suggesting that the velocity of the jet does not vary with plasma temperature: this observation also contradicts the possibility of jets arising from chromospheric evaporation.
However, the plasma emission in the 304 \AA \ is dominated by low-temperature plasma which must originate from the chromosphere initially. One possibility is that chromospheric plasma has already been evaporated after the flare into the corona and in a second step, is ejected with greater velocities during the jet. This scenario can be compatible with events for which the jet occurs with at least a small time delay compared to the flare. However, some of the fast jets in our sample happen before the associated flare. Another scenario explaining these observations would be that the jet originates from the chromosphere directly; however, current models assume that the magnetic reconnection triggering the jet occurs in the corona. 
Finally, we can also assume that chromospheric plasma is available at the base of the jet in the corona due to previous activity in the same active region. 

\subsection{Connection between jets and flares}

This study examined the spatial and temporal link between jets and their associated flares, as well as possible correlations between jet and flare properties.

We found that the flare X-ray emission is generally located at the base of the jet. In a few cases, the thermal X-ray emission extends or is emitted directly from the jet; when detected, the non-thermal emission was emitted from the base of the jet.
The delays observed between the jet and the thermal emission of the flare form a broad distribution with an average close to zero. We did not see any trend suggesting that one triggers the other.
\textbf{This temporal correlation between jets and flares can be compared to the timing between flare-associated CMEs and their corresponding flare: generally, the acceleration phase of the CME is temporally associated with the rise of the SXR flare emission \citep{zhang_2001,maricic_2007,salas-klein_2015}.}
\textbf{A clear temporal correlation has also been observed between the hard X-ray flare emission and the acceleration phase of the CME in a few flare-associated CMEs \citep{qiu_etal_2004, temmer_etal_2010}. In our study, while the jets are in average closely associated in time with the flare emission, the fact that the distribution of delays is broad indicates that the link between the flare and the jet is less evident than for the mentioned CME cases.}
\textbf{The distribution of delays between jets and flares is observed} for both cold plasma emission (304 \AA, $10^{4.7}$ K) and hotter plasma emission (193 \AA, $10^{6.2}$ and $10^{7.3}$ K).
This is in agreement with the interpretation that both jet and flare are the consequence of an episode of magnetic reconnection that involves an open field line, and that the cooler jet velocity reflects the propagation of an Alfvenic perturbation that drives the jet. In this sample, we identified motions in the EUV jets that are compatible with untwisting upflows in 6 jets (helical jets), while 7 jets are clearly straight jets that do not show untwisting motions. Both types of jets have been reproduced by MHD simulations \citep{pariat_etal_2015, pariat_etal_2016}, in which straight jets experience slow interchange reconnection while helical jets are triggered by explosive bursts of interchange reconnection. 
While these observations do not contradict the idea of jets being mini-filament eruptions, we were not able to positively identify mini-filaments in our jet sample \textbf{using the 304 \AA \ emission movies of the jets}. \textbf{In 28} of the studied jets, no indication of a filament detectable with AIA resolution was observed. In the remaining five events, some compact, dark features exist but cannot be convincingly identified as mini-filaments. In addition, we frequently observe multiple surges of plasma in one jet event (see the example presented in figure \ref{time_distance}), which does not seem to support a single ejection of a small-scale filament, but several bursts of ejections driven by Alfvenic perturbations arising from magnetic reconnection episodes.

The distribution of jet-related flare has a probability of 60 \% to be similar to the general distribution of flares already established by previous studies ; this show that jets are probably not preferentially associated with smaller flares. 

We found only a weak correlation (0.14) between the flare SXR peak (GOES peak) and the jet velocity.  This result can be compared with past studies of the relation between the CME speed and the SXR flux of the associated flares. \cite{salas-klein_2015,moon_etal_2002,yashiro_etal_2009} restricted their study to limb CME, for which the projection effects on the velocity as expected to be less important, and found a correlation of 0.48, 0.45 and 0.50 between the CME speed and SXR respectively. \cite{bein_etal_2012,vrsnak_etal_2005} found correlation coefficients of 0.32 and 0.35 for unrestricted events. \cite{moon_etal_2003} found a higher correlation coefficient of 0.93 with a carefully selected set of 8 events for which they corrected the projection effects on the CME velocity. These results seem to indicate that projection effects will significantly lower the correlation between the CME speed and the SXR peak flux of the flare. In our study, we did not correct for projection effects and this may participate to the low correlation. Furthermore, observations of jets seem to show that they can have a high inclination angle from the radial direction, meaning that they rarely propagate radially from the solar surface. Thus, restricting our study to jet events located near the limb would not guaranty a low projection effect on the measured projected velocity. On the other hand, most of the flare considered are B-class or lower, and as discussed in section \ref{discussion:flaresize}, the measured SXR peak flux is more uncertain for flares with class below B7. For example, the flares studied in \cite{salas-klein_2015} are above class B6. It is possible that these uncertainties participate as well in lowering the correlation that we found between the SXR peak flux and the jet projected velocity.
One can note that these conclusions apply to the lack of correlation found between the jet velocity and the flare thermal energy. Several studies have now shown that the thermal energy of flares deduced from isothermal fits to the X-ray spectrum (as done in the present study) can be off by a factor 3 to 5. Moreover, these estimates depends on volumes that are difficult to estimate. For those reasons, the uncertainty on the thermal energy is also significant enough to blur any physical correlation that we could expect. 

Another conclusion could be that the jet velocity simply does not correlates with the flare size or the flare energetics. It is possible that other factors (for example, the magnetic geometry, the magnetic field strength, the amount of magnetic shear...) play a dominant role in the distribution of energy between the flare and the jet, and the velocity of the Alfvenic perturbations that seem to be enlightened by the 304 \AA \ plasma ejection. If this is the case, this would show a fundamental difference between jets and CMEs.

We observed a weak but statistically significant anti-correlation between the jet durations and the SXR peak flux of the flare (background-subtracted GOES peak flux). One tentative explanation is that bigger, more energetic flares are the result of the relaxation of a highly stressed magnetic configuration that will release a lot of energy very fast, in a short time interval. One can imagine that smaller events are produced by quieter and slower episodes of maybe continuous magnetic reconnection. A continuous magnetic reconnection episode will also explain the different bursts sometimes seen in jet events, as discused in section \ref{sub:jetvelo}. This hypothesis could be tested by running MHD simulations of the jet with increasing magnetic stress or shear, for instance, keeping all other quantities constants.

\subsection{Non-thermal energetic electrons in coronal jets}

In our sample, only 1/4 of the events show a non-thermal population of electrons. 
This could be explained by the fact that our sample is mainly composed of small events, for which the non-thermal part of the electron spectrum can easily be below the background signal in the RHESSI detectors. 
Another possibility is that particle acceleration is less efficient in small scale flares, as suggested by the studies of \cite{inglis_christe_2014,warmuth_etal_2016}.

The events for which non-thermal emission was detected do not seem to be associated with the longer or faster jets. When present, the non-thermal emission is not emitted in the jet itself but at its base or in an adjacent loop. This is expected because X-ray intensity is proportional to the plasma density, and therefore the brightest sources of non-thermal X-ray emission are generally located in the chromospheric footpoints of loops/jets. The dynamic range of RHESSI being limited, it is very difficult to observe non-thermal coronal X-ray sources in presence of those footpoint emission sources.

The timing analysis restricted to the events that show non-thermal emission show that in average the jet lightcurve will peak at the same time as the peak of non-thermal emission.

\subsection{Consideration for future studies}

Jets events provide a geometry favoring energetic particle escape from the corona. In this study, we established how often flare-related jets are accompanied by electron acceleration and how the flare and jet energetics relate. However, we did not address how energetic electrons escape along the open magnetic field lines, as this subject is beyond the scope of the paper. A next study should focus on the statistics between type III radio bursts produced by escaping beams of energetic electrons in the corona, to address the question whether jets events are the dominant phenomenon that accompany particle escape. Moreover, only a few studies have looked to jets as sources of impulsive electron events detected in the heliosphere, at 1 A.U. This link should gain new insight with the in-situ data  and radio measurements that will be provided closer to the Sun by Parker Solar Probe and Solar Orbiter.

The detection of X-ray emission in the jet in some cases \citep[see e.g.][]{glesener_etal_2012, bain_fletcher_2009} shows that some energetic electrons are propagating with the hot jet plasma. However, these kind of detection is only rare, due to the limited dynamic range of the past and current X-ray solar telescope. Future instruments based on X-ray focusing optics, demonstrated by the Focusing Optics X-ray Solar Imager (FOXSI) rocket program \citep{krucker_etal_2014,glesener_etal_2016}, should be able to provide imaging and spectroscopy of faint X-ray sources in the solar corona, including in coronal sources and collimated ejections of plasma such as jets. Such observations would provide a new and quantitative insight on the means for energetic particle escape.

\acknowledgments

{

We thank the Lockheed Martin Solar and Astrophysics Laboratory for providing the HEK database that was used for the selection of the events.
We also thank the Helioviewer Project team for providing the Helioviewer web service that was used for the identification of the events.
We thank the RHESSI team for providing the data, the software and the RHESSI flare list that were used for the analysis of the events.
We thank the AIA team for providing the data, routines and the cutout service to access and process the AIA data.
We thank the GOES team for providing the SXR flux data.
This work is supported by the 2018 ISSI Team ``Solar flare acceleration signatures and their connection to solar energetic particles''. 
MJ was supported by the Undergraduate Research Opportunities Program of the University of Minnesota. 
}


\appendix

\section{Appendix: list of EUV jets} \label{sec:listjets}

\begin{small}
\begin{longtable}{llllllllllllll}
\hline
\multicolumn{6}{c}{Jet}                                                                        & \multicolumn{7}{c}{Flare}                                                                          \\
Date       & $t_{s,jet}^{304} $ & $t_{p,jet}^{304} $ & Dur. & \multicolumn{2}{c}{Position} &  AR  & Flare num. & \multicolumn{2}{c}{Position} & Peak time & class & $E_{th}$ & $\delta$ \\
\hline
08/02/2010  & 17:18:26 & 17:27:02  & 18.8  & -153 &  117 & 11092 & 10080216  & -125 &  114 & 17:24:30 & A8.2 &  1.2 &     \\
03/29/2011  & 20:18:56 & 20:28:08  & 24.2  &  242 & -204 & 11176 & 11032921  &  259 & -240 & 20:53:22 & A4.4 &  1.4 &     \\
04/01/2011  & 03:54:44 & 04:00:20  & 20.8  &  702 & -234 & 11176 & 11040104  &  691 & -225 & 03:53:42 & C2.8 &  5.2 & 4.3 \\
12/11/2011  & 03:22:20 & 03:25:56  &  8.2  & -565 & -299 & 11374 & 11121105  & -627 & -269 & 04:10:50 & A9.0 &  0.6 &     \\
12/11/2011  & 12:18:56 & 12:26:56  & 32.0  & -487 & -322 & 11374 & 11121110  & -506 & -305 & 12:05:38 & A3.3 &  1.0 &     \\
06/30/2012  & 18:26:32 & 18:30:48  & 13.0  & -287 &  226 & 11513 & 12063050  & -290 &  205 & 18:31:50 & M1.5 & 10.1 & 3.6 \\
09/12/2012  & 04:19:07 & 04:24:43  &  7.0  &  934 & -204 & 11584 & 12091248  &  936 & -203 & 03:56:54 & A4.0 &  0.2 &     \\
09/12/2012  & 04:27:43 & 04:37:43  & 27.0  &  940 & -214 & 11584 & 12091249  &  940 & -212 & 04:51:18 & A8.2 &  1.2 &     \\
10/10/2012  & 14:29:07 & 14:30:43  & 20.8  &  577 & -336 & 11585 & 12101032  &  550 & -358 & 14:30:54 & C1.8 &  3.1 & 3.3 \\
10/19/2012  & 18:07:07 & 18:09:31  & 19.4  & -938 & -259 & 11598 & 12101919  & -943 & -235 & 18:08:50 & B3.1 &  0.4 &     \\
04/24/2013  & 12:02:31 & 12:04:55  &  7.8  &  114 &  482 & 11727 & 13042449  &  113 &  480 & 12:16:58 & B1.8 &  0.2 &     \\
04/24/2013  & 12:13:43 & 12:16:19  & 10.2  &  100 &  480 & 11727 & 13042449  &  113 &  480 & 12:16:58 & B1.8 &  0.2 &     \\
04/28/2013  & 20:55:19 & 21:00:55  & 18.8  & -330 &  202 & 11731 & 13042858  & -310 &  209 & 21:06:06 & B4.9 &  3.3 &     \\
05/04/2013  & 23:16:07 & 23:20:31  & 31.2  & -114 & -297 & 11734 & 13050454  & -65  & -297 & 23:31:46 & B8.0 &  3.2 &     \\
06/17/2013  & 08:46:31 & 08:49:43  & 19.6  & -279 & -394 & 11769 & 13061729  & -303 & -389 & 08:52:30 & B1.1 &  4.2 &     \\
06/18/2013  & 15:14:43 & 15:17:19  & 14.6  &  -45 & -422 & 11772 & 13061834  &  -46 & -423 & 15:21:06 & B3.8 &  0.3 &     \\
10/29/2013  & 03:05:43 & 03:07:31  & 12.6  &  970 &  143 & 11875 & 13102913  &  965 &  150 & 03:13:42 & C1.4 &  1.5 & 3.4 \\
10/29/2013  & 03:30:31 & 03:30:19  &  6.0  &  971 &  137 & 11875 & 13102916  &  961 &  131 & 03:36:18 & C1.3 &  0.5 & 2.1 \\
12/23/2013  & 17:34:07 & 17:38:43  & 14.4  &  834 & -301 & 11928 & 13122361  &  875 & -291 & 17:39:34 & B3.7 &  0.8 &     \\
04/10/2014  & 21:13:19 & 21:24:43  & 23.0  & -936 & -235 & 12035 & 14041026  &      &      & 21:01:02 & B2.9 &      &     \\
04/10/2014  & 21:33:19 & 21:47:43  & 51.0  & -928 & -278 & 12035 & 14041027  &      &      & 22:58:38 & B3.1 &      & 3.8 \\
04/10/2014  & 23:31:19 & 23:38:55  & 23.0  & -928 & -246 & 12035 & 14041031  & -950 & -322 & 23:25:30 & A4.9 &  5.0 &     \\
04/10/2014  & 22:58:55 & 23:12:19  & 37.4  & -908 & -319 & 12035 & 14041029  & -939 & -322 & 23:16:54 & A1.2 &  4.7 &     \\
04/11/2014  & 00:35:07 & 01:00:07  & 70.2  & -903 & -324 & 12035 & 14041102  & -920 & -290 & 00:44:06 & B6.7 &  1.7 & 3.8 \\
04/11/2014  & 01:36:43 & 01:41:43  & 44.6  & -936 & -294 & 12035 & 14041108  & -926 & -311 & 02:35:42 & B1.3 &  2.0 &     \\
10/04/2014  & 16:57:19 & 17:01:31  &  8.8  &  464 & -306 & 12181 & 14100413  &  460 & -307 & 17:03:30 & B3.5 &  0.1 &     \\
11/13/2014  & 17:15:55 & 17:18:31  & 15.4  & -914 & -262 & 12209 & 14111357  & -977 & -305 & 17:23:18 & B2.0 &  1.3 &     \\
11/22/2014  & 06:00:55 & 06:03:07  &  4.2  &  368 & -253 & 12209 & 14112210  &  437 & -262 & 06:03:30 & C5.7 &  6.1 & 1.8 \\
11/22/2014  & 06:12:43 & 06:16:55  & 12.8  &  371 & -256 & 12209 & 14112212  &  384 & -267 & 06:17:46 & C2.3 &  1.9 &     \\
03/09/2015  & 01:10:43 & 01:14:07  & 14.4  & -765 & -211 & 12297 & 15030901  & -774 & -201 & 01:13:18 & C1.3 &  1.9 &     \\
08/29/2015  & 17:01:54 & 17:04:54  &  4.4  &  912 & -280 & 12403 & 15082926  &  907 & -278 & 17:07:38 & B2.4 &  0.3 & 4.2 \\
12/19/2015  & 02:00:06 & 02:02:42  &  8.2  &  564 & -292 & 12468 & 15121903  &  584 & -285 & 02:02:46 & C2.7 &  1.0 &     \\
09/27/2016  & 20:15:42 & 20:23:42  & 21.6  &  695 & -316 & 12597 & 16092710  &  723 & -317 & 20:15:30 & A6.3 &  1.5 &     \\
\hline
\caption{List of EUV jets considered in this paper. $t_{s,jet}^{304} $ and $t_{p,jet}^{304}$ are the start and peak times of the jets deduced from 304 \AA \ data analysis. Durations (dur.) are in minutes, positions are in arcseconds. $T.$ is the type of the jet: blowout (bl.), standard (st.) or uncertain (un.). AR is the active region number. The flare number and flare position are from the RHESSI flare list, the class is background-subtracted GOES class calculated with the 1-8 \AA \ lightcurve. $E_{th}$ is the thermal energy of the flare deduced from RHESSI, in $10^{28}$ ergs, $\delta$ is the spectral index of the non-thermal component of the RHESSI spectrum.}
\label{listofjets}
\end{longtable}
\end{small}

\section{Appendix: volume estimates from RHESSI imaging} \label{sec:volumes}


Calculating the volume from RHESSI imaging can be difficult, because of the partial Fourier coverage of the instrument. Different imaging algorithms will give different estimates, and the CLEAN algorithm is known to overestimate the source size \citep{aschwanden_etal_2004, schmahl_et_al_2007, kontar_et_al_2010}. In this study, we used the estimates derived from the visibility forward fit algorithm, which provide a measure of the uncertainty of the size of the source.  This is described in section \ref{sec:imaging}. Nevertheless, we compared the images produced with the visibility forward fit to the CLEAN image to verify the coherence of the volume estimates. The CLEAN images were produced with a beam factor of 1.2, this value being arbitrary chosen. The surface of the source is calculated as the surface $S_{clean}$ covered by the 50 \% contour of the CLEAN image. The CLEAN volume is calculated with the assumption of a spherical source:
\begin{equation}
V_{clean} =  \frac{4 \sqrt{\pi}}{3} S_{clean}^{3/2}
\end{equation}
The error on that volume is estimated to be 50 \% of the value.

For most of the flares, the CLEAN volume is greater than the visibility forward fit volume, as expected. However, there is a good correlation between the two volumes (0.89), which confirms that the volumes used in this paper are representative of the X-ray source sizes.

\section{Appendix: AIA and RHESSI images of the jets} \label{sec:alljets}

\begin{figure}
\includegraphics[width=0.99\linewidth]{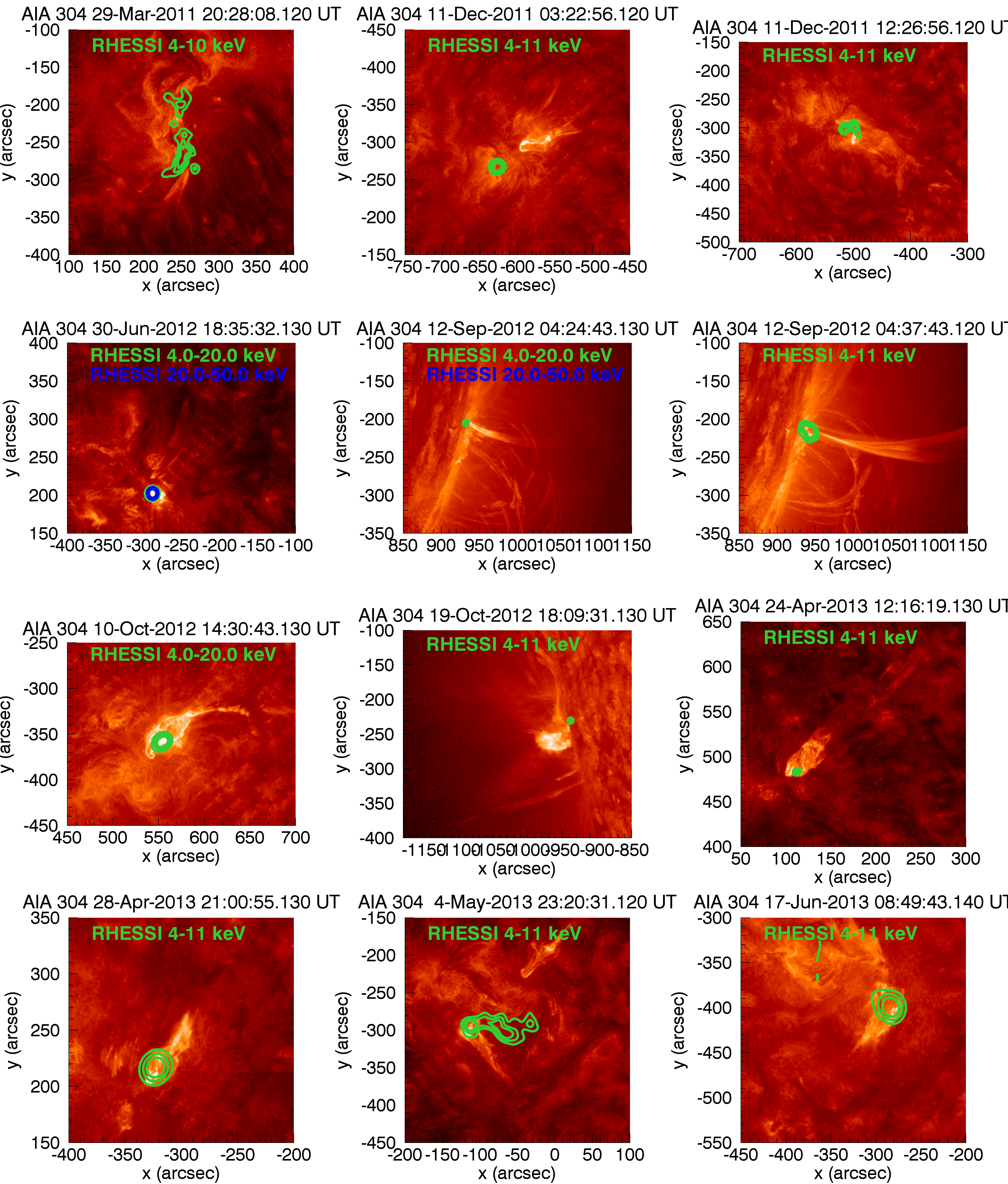}
\label{alljets1}
\end{figure}

\begin{figure}
\includegraphics[width=0.99\linewidth]{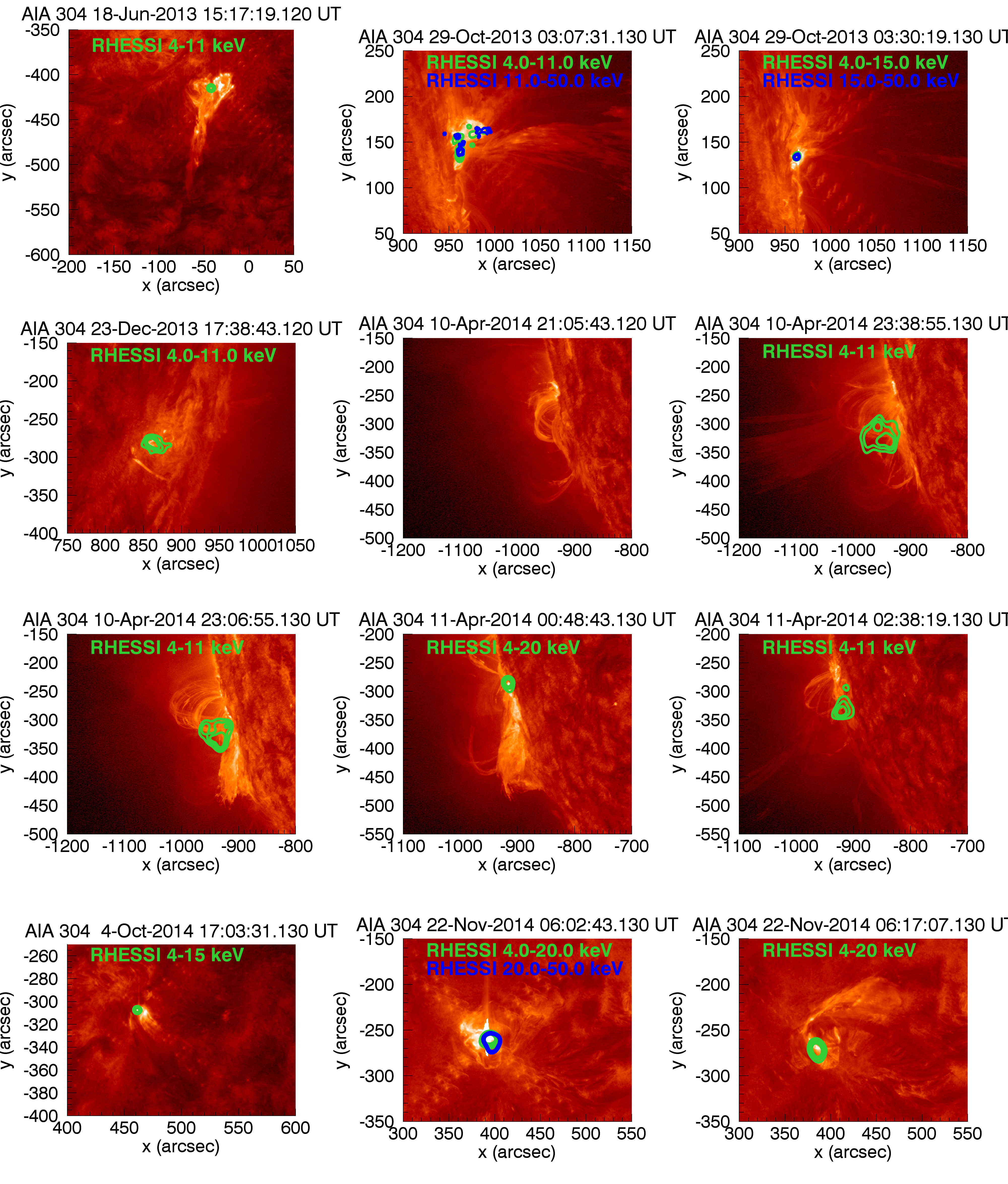}
\label{alljets2}
\end{figure}

\begin{figure}
\includegraphics[width=0.99\linewidth]{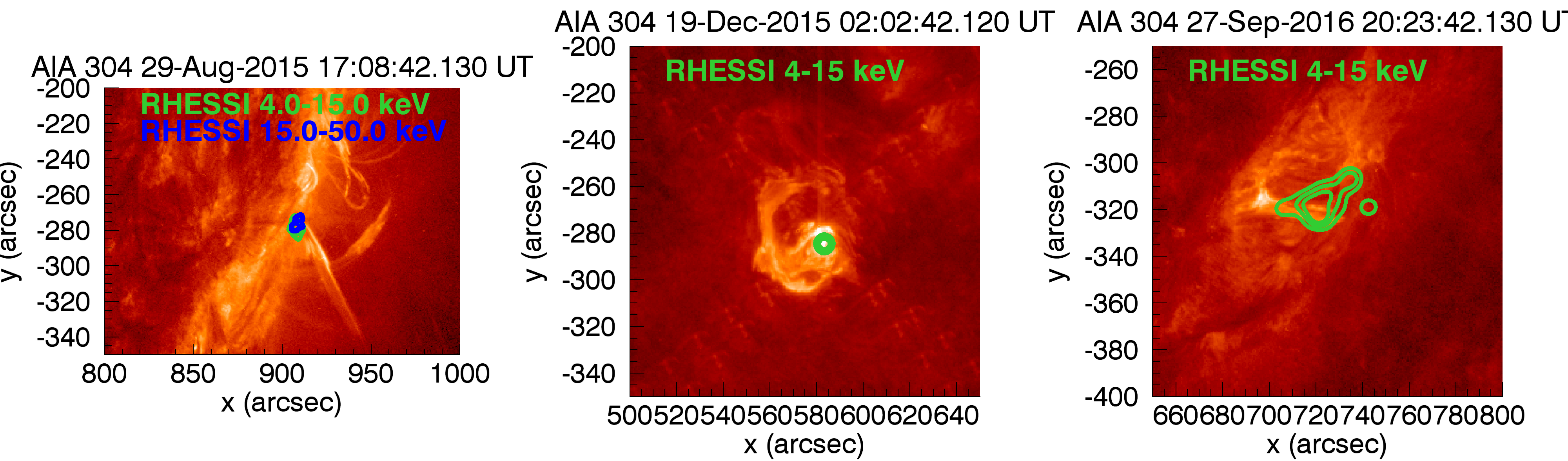}
\caption{AIA 304 \AA \ images of the jets and contours of the RHESSI emission (thermal in green and non-thermal in blue when present).}
\label{alljets3}
\end{figure}


\bibliographystyle{aasjournal} 
\bibliography{zmabiblio} 




\end{document}